\documentclass[twoside,leqno,twocolumn]{article}
\usepackage[letterpaper]{geometry}

\usepackage{siamproceedingsmacros_022425/siamproceedings}

\Crefname{ALC@unique}{Line}{Lines}
\newcounter{myalg}
\AtBeginEnvironment{algorithmic}{\refstepcounter{myalg}}
\makeatletter
\@addtoreset{ALC@unique}{myalg}
\makeatother

\usepackage{lscape}
\usepackage{afterpage}
\usepackage{amsopn}

\newcommand{\amoc}{{\footnotesize $\leq$}1}
\newcommand{\aloc}{{\footnotesize $\geq$}1}

\newcommand{\Procedure}[2]{%
  \STATE \textbf{Procedure} #1(#2)%
  \begin{ALC@g} 
}
\newcommand{\EndProcedure}{%
  \end{ALC@g}%
  \STATE \textbf{End Procedure}%
}

\newcommand\relatedversion{}

\renewcommand\relatedversion{\thanks{Related repository with the supplementary materials: \protect\url{https://doi.org/10.5281/zenodo.17344937}}}

\usepackage[mathscr]{eucal}
\usepackage{subcaption}

\newcommand{\bigO}{\mathcal{O}}

\usepackage{comment,color}

\usepackage{mathtools}

\newcommand{\rem}[3]{\textcolor{blue}{\textsc{#1 #2:}}
  \textcolor{red}{\textsf{#3}}}

\renewcommand{\rem}[3]{}

\newcommand{\FOP}[1]{$\mathrm{FOP}_k$}

\makeatletter
\renewcommand\paragraph{%
  \@startsection{paragraph}
    {4}
    {\z@}
    {3.25ex \@plus1ex \@minus.2ex}
    {-1em}
    {\normalfont\normalsize\bfseries\addperiod}}
\newcommand{\addperiod}[1]{#1\@addpunct{.}}
\usepackage{csquotes}

\usepackage[noend]{algorithmic}

\AtBeginDocument{%
  \@ifpackageloaded{algorithm2e}{%
    \SetArgSty{upshape}
    \SetKwProg{Fn}{Function}{}{}
  }{}
}
\title{Engineering Dominating Patterns: A Fine-grained Case Study\relatedversion}
\author{Jonathan Dransfeld\footnotemark[2] \and Marvin K\"unnemann\footnotemark[2] \and Mirza Redzic\thanks{Karlsruhe Institute of Technology. Research supported by the Deutsche Forschungsgemeinschaft (DFG, German Research Foundation) – 462679611.} \and Marcus Wunderlich\thanks{University of Augsburg}}

\begin{document}
\fancyfoot[R]{\scriptsize{Copyright \textcopyright\ 2026 by SIAM\\
Unauthorized reproduction of this article is prohibited}}

\date{}

\maketitle
\begin{abstract}
    The \emph{Dominating $H$-Pattern} problem generalizes the classical $k$-Dominating Set problem: for a fixed \emph{pattern} $H$ and a given graph $G$, the goal is to find an induced subgraph $S$ of $G$ such that (1) $S$ is isomorphic to $H$, and (2) $S$ forms a dominating set in $G$. Fine-grained complexity results show that on worst-case inputs, any significant improvement over the naive brute-force algorithm is unlikely, as this would refute the Strong Exponential Time Hypothesis. Nevertheless, a recent work by Dransfeld et al. (ESA 2025) reveals some significant improvement potential particularly in \emph{sparse} graphs. 
    
    We ask: Can algorithms with conditionally almost-optimal worst-case performance solve the Dominating $H$-Pattern, for selected patterns $H$, efficiently on practical inputs? We develop and experimentally evaluate several approaches on a large benchmark of diverse datasets, including baseline approaches using the Glasgow Subgraph Solver (GSS), the SAT solver Kissat, and the ILP solver Gurobi.
    Notably, while a straightforward implementation of the algorithms -- with  conditionally close-to-optimal worst-case guarantee  --  performs comparably to existing solvers, we propose a tailored Branch-\&-Bound approach -- supplemented with careful pruning techniques -- that achieves improvements of up to two orders of magnitude on our test instances.
\end{abstract}

\section{Introduction}
The \emph{$k$-Dominating Set} problem is one of the most extensively studied problems in graph theory. Given a graph $G=(V,E)$, the goal is to find a subset of vertices $S\subseteq V$ of size $k$, such that every vertex is either (1) contained in $S$, or (2) adjacent to some vertex $v\in S$.
This problem is often considered to be one of the central pillars of two different fields.
\begin{itemize}
    \item In Graph Theory and Discrete Mathematics: it gives rise to some of the natural structural parameters of a graph (e.g. \emph{domination number}, \emph{bondage number})~\cite{FuruyaM15,KralSV12,MacGillivrayS96,RadV11}. Moreover, the concept of dominating set is linked to some classical conjectures in the field such as the \emph{Vizing conjecture}~\cite{AharoniS09,BresarR09,Zerbib19}. 
    \item In Algorithms and Complexity Theory: $k$-Dominating Set is one of the classic NP-complete problems (when $k$ is a part of the input). In Parameterized Complexity Theory, it is often regarded as the most natural $W[2]$-complete problem (when parameterized by the solution size) \cite{DowneyF95}; in Fine-Grained Complexity Theory, it is one of the first problems for which the tight conditional lower bounds under SETH have been established \cite{EisenbrandG04,PatrascuW10}. Dominating Set problem was also studied from the lens of the approximation algorithms \cite{CKLM17,SolomonU23,SLM19}.
\end{itemize}
Beyond the study in theoretical research, this problem appears naturally in many practical applications (e.g. facility location problems \cite{CorcoranG21,FrankiewiczGGLM20}, network design \cite{SouL17,SmithH22}, etc.). 
However, in order to model many practical problems, using the simple $k$-Dominating Set problem is not enough. This motivated a plethora of related variants of \emph{domination problems} to be studied, where in general we are given a graph $G$ and are looking for a set $S$ consisting of $k$ vertices in $G$ that (1) form a dominating set, \emph{and} (2) satisfy some additional internal property (e.g. induce a connected subgraph).
Some notable well-studied examples of such domination problems include \emph{Connected Dominating Set} ($S$ induces a connected subgraph of $G$)~\cite{DraganicK25,FominLST12}, \emph{Independent Dominating Set}\footnote{Equivalent to the Maximal Independent Set problem.} ($S$ induces an independent set in $G$)~\cite{Byskov03,Eppstein05}, \emph{Total Dominating Set} ($S$ induces a subgraph with no isolated vertices)~\cite{HenningY09,PetrPV24}, and many more.
An important class of domination problems is the class of \emph{Dominating Patterns}: given a graph $G$, decide if there exists a dominating set $S$ that dominates $G$ and induces a graph isomorphic to $H$ for a $k$-vertex (pattern) graph $H$.\footnote{For some notions, non-induced patterns are studied as well (e.g., Dominating Cycle or Matching Domination, Dominating Path). However, in this work, unless stated otherwise, we always consider \emph{induced} patterns.}  
For various choices of $H$, (e.g. Cycles~\cite{BondyF87,KeilS92}, Paths~\cite{FaudreeGJW17,FaudreeFGHJ18}, Cliques~\cite{BourgeoisCEP12,CozzensK90}) this problem has been heavily studied in the field of Graph Theory.
A first systematic study of such problems from the perspective of fine-grained algorithms was done recently by Künnemann and Redzic \cite{KunnemannR24},
and extended further in the follow-up work by Dransfeld, Künnemann and Redzic \cite{DransfeldKR25}. 
They first observe that in dense graphs, this problem requires $n^{|V(H)|- o(1)}$ time in worst-case (unless the standard $k$-OV complexity conjecture fails), with a fast-matrix-multiplication-based algorithm matching the lower bound for sufficiently large patterns.
However, this conditional lower bound intrinsically applies only to dense graphs. Since many graph classes of theoretical interest (e.g. planar graphs, bounded treewidth graphs) and many real-world graphs (e.g. road networks, social networks) are sparse, this motivates a more detailed analysis of the problem's complexity in sparse graphs.
Indeed, recent work~\cite{DransfeldKR25} shows that, for any fixed pattern $H$ of size $k\geq 2$ (with the exception of $K_3$), the problem has a fine-grained time complexity of essentially $n^{\rho(H)}m^{\frac{k-\rho(H)}{2}+o(1)}$, where $\rho(H)$ is a parameter that only depends on the pattern graph $H$.

Even though the theoretical complexity of detecting a dominating $H$-pattern is well understood, there is still a gap between this theoretical understanding and practical implementations.
Particularly, given the interest in Dominating $H$-Patterns from the perspective of structural graph theory, we ideally wish to retrieve, from a given large database (e.g. House of Graphs~\cite{house-of-graphs}), all graphs exhibiting a dominating $H$-pattern. We thus~ask:


\begin{enumerate}
\item Can we implement algorithms with close-to-optimal worst-case complexity that also work well on many realistic inputs?
\item Do fast algorithms for Pattern Domination inherently require a combination of aspects of Pattern Detection and Dominating Set, as suggested by theoretical bounds?
\end{enumerate}
To elaborate on the second aspect: the Fine-grained Complexity analysis reveals that Pattern Domination combines central aspects of Pattern Enumeration and Dominating Set. For one,  the fastest algorithms given in \cite{DransfeldKR25} work as follows: we efficiently find (a representation of) relevant candidate occurrences of the pattern $H$ in $G$, and check whether at least one of them dominates the graph. In fact, for some patterns, it suffices to look at only a $\Theta(m/n^2)$-fraction of worst-case occurrences (saving a linear factor if $m=\Theta(n)$). On the other hand, one can show that this approach is essentially conditionally optimal, via reductions that adapt a well-known SETH-based lower bound for the Dominating Set problem~\cite{PatrascuW10}, see~\cite{DransfeldKR25}.

\paragraph{Our Contribution} As our main contribution, we develop a fast, light-weight implementation of a Dominating $H$-Pattern solver for selected patterns $H$ (cycles $C_k$, paths $P_k$ and induced matchings $M_k$ of various sizes $k$) that significantly outperforms all the natural baseline solvers we tested against.
Our pruning strategy crucially exploits an interplay between the two main aspects of the problem -- domination and subgraph isomorphism --  in order to prune the branches and thus heavily reduce the search space for the considered (realistic) instances. 

Furthermore, we explore the use of the fine-grained reduction in the algorithm engineering process.
Here, we particularly exploit known reductions for generating hard, but insightful benchmark sets. Specifically we generate seemingly
\emph{hard} $k$-OV instances via a carefully chosen randomized approach. This allows us to construct (conditionally) hard instances of the Dominating $H$-Pattern problem, for which intuitively, any solver needs to look at a large number of candidate solutions (no significant pruning is possible, unless SETH fails). In our experiments, these instances serve as models for worst-case inputs.
\paragraph{Further Related Work}
The practical relevance of domination-type problems has recently been further highlighted by the PACE 2025 (Parameterized Algorithms and Computational Experiments) Challenge, which features Dominating Set as one of its core problems.
While the challenge shares thematic overlap with our work, the results of PACE 2025 are still under evaluation at the time of writing. Consequently, our work was developed independently and could not be influenced by any outcomes or insights from this year's competition.

\section{Methods and Approaches}

Our starting point is to implement simplified versions of the algorithms given in~\cite{KunnemannR24,DransfeldKR25} towards obtaining close-to-optimal worst-case performance. These algorithms essentially enumerate all occurrences of the pattern $H$ containing at least one node of sufficiently large degree (see below for a discussion) and perform a simple domination check for each such occurrence. Our experimental evaluations (which focus on the natural patterns of matchings, cycles and paths) reveal that for several benchmarks, this simple approach will not outperform other baseline solutions (in particular generic SAT/ILP solvers, see Section~\ref{ssec:baseline-approaches}). Instead, we employ additional heuristics (Section~\ref{ssec:heuristics}), and more importantly, design a Branch-\&-Bound approach that exploits both main aspects of the problem (domination and subgraph isomorphism) simultaneously to reduce the branching space (see Sections~\ref{sssec:proto-bnb}-\ref{sssec:bnb-paths}).



\subsection{Heuristics}\label{ssec:heuristics}
We first describe heuristics that will be beneficial for several approaches.
\paragraph{Low-degree-dominator start}
Let $\delta$ be the minimum degree of $G$ and $v$ be a vertex of degree $\delta$. 
Any potential solution forms a dominating set, and hence must either (1) contain $v$, or (2) contain a vertex $u$ that is adjacent to $v$.
Hence, in order to find our first candidate solution vertex, it suffices to look at the closed neighborhood $N[v] := \{ u \mid \{u,v\} \in E(G)\}\cup \{v\}$ of $v$.
Recall that in graphs with $n$ vertices and $m$ edges, the minimum degree $\delta$ is at most $\frac{2m}{n} = \bigO(\frac{m}{n})$.
Note that in previous work~\cite{FischerKR24,KunnemannR24,DransfeldKR25}, the same asymptotic improvement is obtained by noticing that any $k$-dominating set contains a vertex of degree at least $\left(\frac{n}{k}-1\right)$. However, our experiments indicate that using the min-degree heuristic is significantly better on most test instances.
Furthermore, as we shall exploit later, this heuristic can be applied repeatedly, by marking each vertex either \emph{dominated} (if it is in the closed neighborhood of a guessed vertex), or  \emph{non-dominated}, and then finding a vertex $v'$ of minimum degree among the non-dominated vertices. Again, any dominating $H$-pattern must include a vertex of $N[v']$. This will give rise to a more sophisticated approach detailed below.


\paragraph{$3$-scattered set lower bound}
A \emph{$3$-scattered set} of a graph $G=(V,E)$ is a vertex subset $S \subseteq V$ such that all vertices in $S$ have a pairwise hop-distance of at least $3$ \cite{3ScatteredSet}. The size of any $3$-scattered set is a natural lower bound of the domination number of a graph, as no vertex $v \in V$ can dominate more than one vertex in $S$. 
We exploit this observation in several of our solvers to prune those branches that do not satisfy this lower bound, and hence cannot produce a valid solution.

More formally, during the preprocessing step, we first greedily compute a $3$-scattered set $S$ of our input graph. 
Let $P$ be a partial solution of size $\ell$, and let $S_{P}:= S\setminus N[P]$ be the set of vertices in $S$ that are not dominated by $P$.
By the argument above, if $|S_P|> k-\ell$, then there exists no dominating set $D$ of size $k$ that contains $P$.
Hence, for any pattern $H$ containing at most $k$ vertices, we can prune any branch that yields $P$ as a partial solution.

\subsection{Baseline approaches}\label{ssec:baseline-approaches}

As baselines, we investigate two approaches: (1) Adapting an established solver for subgraph isomorphism to also check domination. To obtain a fair comparison, we also incorporate the heuristics developed for our novel solvers. (2) Encoding the Dominating $H$-Pattern problem as SAT or ILP instances. These two baseline approaches roughly correspond to approaching the problem either from the Pattern Detection angle or the Dominating Set angle. We remark that for dominating-set-type problems, particularly for the generalization Hitting Set, previous works suggest that the fastest solutions are obtained via ILP and specialized Branch-\&-Bound methods, see, e.g.~\cite{Blasius0SW22}.

\paragraph{Pattern Enumeration}
To solve the Dominating $H$-Pattern problem from the pattern detection angle, our baseline approach is to enumerate all occurrences of the pattern $H$ in the host graph $G$ and check, for each of them, if they dominate $G$ entirely. For our experiments, we chose the Glasgow-subgraph-solver (short GSS) by McCreesh, Prosser \& Trimble \cite{Gss20} for $H$-enumeration. This is a state-of-the-art solver for subgraph isomorphism problem including (induced) pattern enumeration that uses constraint programming, and is easily adaptable to new variants. 
To check for a given occurrence of $H$ if it dominates $G$, we use a precalculated adjacency matrix (for inclusive neighborhoods) and check if the OR-ed rows of the assigned target vertices contain any $0$ entry. 
For fair comparison, we also incorporated all other heuristics and techniques that we use in our Branch-\&-Bound solvers in the GSS, which is possible since the GSS also follows the Branch-\&-Bound paradigm. Therefore, the Low-degree-dominator start and the $3$-scattered-set heuristic (explained in \Cref{ssec:heuristics}) are applicable here without any significant modifications. We refer to the adapted version of the GSS as \emph{enhanced GSS (eGSS)}.

\paragraph{Generic Solvers}
As many other optimization problems, the dominating $H$-pattern problem can be written as a SAT-formula, and thus also as an Integer Linear Program (ILP). For both formulations we define a variable $x_{v,p}$ for every element in $(v,p) \in V(G)\times V(H)$, each of which is set to \texttt{true} (for SAT) or $1$ (for ILP) if and only if $v \in V(G)$ is assigned to $p \in V(H)$. In the following, we describe a generic model for the dominating $H$-pattern problem with the help of so-called \amoc- and \aloc-constraints. Given a set of variables $X$, an \amoc-constraint (resp. \aloc-constraint) on $X$ enforces that at most one (resp. at least one) variable in $X$ is set to \textsc{true} (i.e. $1$ in case of ILP's). The \aloc-constraints are easy to model as both SAT, as well as the ILP constraints. However, while \amoc-constraints are also easy for ILP, it is harder to model them in a SAT-formula. We describe different variants to model \amoc-constraints in SAT after the definition of the model.

To make sure that the solution of our SAT/ILP formulations is indeed a valid solution for the dominating $H$-pattern instance, we first have to ensure that exactly one vertex $v \in V(G)$ is assigned to every fixed $p \in V(H)$. This is done by an \amoc- and an \aloc-constraint on $\{x_{v,p} \mid v \in V(G)\}$ for every pattern vertex $p \in V(G)$. Next, the found solution has to be dominating. Thus, for every target vertex $v \in V(G)$, we add an \aloc-constraint for $N_G[v] \times V(H)$. To ensure that the solution actually induces $H$, we forbid contradicting assignments, i.e., set variables $x_{p,v}$ and $x_{q,u}$ such that $\{p,q\} \in E(H)$ but $\{v,u\} \notin E(G)$ or vice versa. This is done by two \amoc-constraints on $\{x_{p,v},x_{q,u}\}$ and $\{x_{p,u},x_{q,v}\}$ respectively for every combination of (1) pattern edge $\{p,q\} \in E(H)$ and host non-edge $\{u,v\} \notin E(G)$ and (2) every combination of pattern non-edge $\{p,q\} \notin E(H)$ and host edge $\{u,v\} \in E(G)$.

We use two different encoding for \amoc-constraints for SAT on a set of variables $X$. The first one explicitly states that among every pair $x,y \in X$ with $x\neq y$ not both variables may be set to true. Thus, this encoding needs $\mathcal{O}(|X|^2)$ clauses for an \amoc-constraint on $X$. In contrast to that, we use the \emph{ladder-encoding} introduced by H{\"o}lldobler \& Nguyen, which only needs linearly many clauses to enforce the same constraint \cite{holldobler13ladder}. However, this comes at the cost of linearly many additional variables and unit propagation to enforce the constraint. We refer to algorithms using the ladder-encoding for \amoc-constraints together with a SAT-solver as \emph{ladder-SAT} while the other ones are called \emph{(default) SAT}.

For our experiments we use the \emph{kissat}-SAT-solver by Biere et al. \cite{kissat24}, one of the best performing SAT-solver at the 2024 SAT-competition, to solve our SAT-formulas. For the ILP-formulation, we use the \emph{Gurobi} as a state-of-the-art solver \cite{gurobi}.

\subsection{Integrated approaches}\label{ssec:bnb}

\noindent
In this subsection, we describe our Branch-\&-Bound approach for the Dominating $H$-Pattern problem. We first describe the general Branch-\&-Bound framework that all our solvers follow. This framework and pseudocode for the main recursive function are shown in \Cref{alg:bnb-start,alg:bnb-rec} respectively.
Based on that, we then explain how this framework can be realized for general patterns (see \Cref{sssec:proto-bnb}) as well as for specific patterns like matchings (\Cref{sssec:bnb-matchings}), cycles (\Cref{sssec:bnb-cycles}) and paths (\Cref{sssec:bnb-paths}).

Before we start with description of the Branch-\&-Bound framework, we fix some notations. Along the process of all Branch-\&-Bound solvers, we store the current assignments in a \emph{partial assignment}, which is a partial function $\sigma{\colon}V(H){\to}V(G)$ from the vertices of the pattern graph $H$ to the target vertices $V(G)$, where we denote an undefined assignment for a pattern vertex $v$ as $\sigma(v)=\bot$. The way a partial solution stores its assignments and handles queries highly depends on the specific patterns. For that we refer to the respective sections of the specific solvers. 

In every step of the recursion, the current partial assignment $\sigma$ can be extended by a set of \emph{candidates}. A \emph{candidate} is a pair $(p,v) \in V(H) \times V(G)$ containing a pattern vertex $p$ and a target vertex $v$ that can be assigned to $p$ based on the already existing assignments in $\sigma$. 
The rest of this subsection is dedicated to describing the general framework of our Branch-\&-Bound solvers, and providing more details on our recursive function.

\paragraph{Initialization and Framework}
All Branch-\&-Bound solvers start by defining an empty partial assignment $\sigma$ (\Cref{line:bnb-start:partial-solution} of \Cref{alg:bnb-start}). After that, we get a first set of candidates to extend the initially empty assignment. This is done by the function \textsc{InitialCandidates($G$)} in \Cref{line:bnb-start:initCands}. Again, the set of candidates we start with depends on whether we have to deal with arbitrary pattern or whether we know the pattern in advance and use given properties of them. After that, we start the recursive algorithm by calling the main function \textsc{extendSolution()}, given the initial candidate set $C_0$ and an empty partial assignment $\sigma$, 


\begin{algorithm}
\caption{Branch\,\&\,Bound-Solver Framework}
\label{alg:bnb-start}
\hspace*{\algorithmicindent} \textbf{Input:} Host graph $G$, pattern graph $P$\\
\hspace*{\algorithmicindent} \textbf{Output:} dominating assignment $\sigma{\colon}V(P){\to}V(G)$\\[-12pt]
\begin{algorithmic}[1]
\STATE{Set $\sigma(u)  \leftarrow \bot$ for every $u\in V(P)$\label{line:bnb-start:partial-solution}}
\STATE{$C_0 \gets $ \textsc{InitialCandidates}$(G)$}\label{line:bnb-start:initCands}\;
\RETURN \textsc{extendSolution}($\sigma$, $C_0$)\label{line:bnb-start:start-rec}\;
\end{algorithmic}
\end{algorithm}


\begin{algorithm}
\caption{Recursive Function \textsc{extendSolution}}
\label{alg:bnb-rec}
\begin{algorithmic}[1]
    \Procedure{\textsc{extendSolution}}{$\sigma, C$}
        \IF{$\sigma$ is complete\label{line:bnb-rec:complete}}
            \IF{$\sigma$ dominates $G$}\label{line:bnb-rec:domcheck} 
                \RETURN $\sigma$
            \ELSE
                \RETURN FAIL
            \ENDIF
        \ENDIF
        \IF{\textsc{Bound}($\sigma$)\label{line:bnb-rec:bound}}
            \RETURN FAIL
        \ENDIF
        \FOR{$(u,v) \in C$\label{line:bnb-rec:loop}}
            \STATE{update $\sigma(u) \gets v$ \label{line:bnb-rec:assign}}
            \STATE{$C'$ $\leftarrow$ \textsc{ExtensionCandidates}($\sigma$, $G$)\label{line:bnb-rec:guesses}}
            \STATE $\sigma' \gets $\textsc{extendSolution}($\sigma$, $C'$)
            \IF{$\sigma' \ne \mathrm{FAIL}$ \label{line:bnb-rec:recurse}}  
                \RETURN $\sigma'$
            \ENDIF
            \STATE{update $\sigma(u) \gets \bot$}
        \ENDFOR
    \EndProcedure
\end{algorithmic}
\end{algorithm}

\paragraph{Our Recursive Process} The function \textsc{extendSolution()} (see \Cref{alg:bnb-rec}) implements the actual Branch-\&-Bound idea of the solvers.  This function takes a partial assignment $\sigma{\colon}V(H){\to}V(G)$ as well as a list of candidates $C\subseteq V(H)\times V(G)$. Those candidates come from a \emph{safe branching rule}, meaning if $\sigma$ can be extended to a valid complete solution, then at least one of the candidates $c \in C$ is part of at least one of these valid complete solutions based on $\sigma$. These branching rules depend on the specific pattern.

In each recursive call, \textsc{extendSolution()} first checks whether the given partial assignment is complete, i.e., every pattern vertex has an assignment. If so, we check whether the host vertices chosen by $\sigma$  dominate the host graph $G$. If this is the case, $\sigma$ is returned as a valid solution. Otherwise, $\sigma$ cannot be extended to a proper solution, hence we can return FAIL. In every case, where the partial assignment $\sigma$ is not yet completed, we are looking for ways to extend $\sigma$ to a proper solution. If there are techniques or heuristics to recognize that the current assignment $\sigma$ cannot be extended to a proper solution, we use them in the \textsc{Bound}-check in \Cref{line:bnb-rec:bound}. When there is no other way to bound this branch, we iterate through the list of candidates and try to extend $\sigma$ with them (starting in \Cref{line:bnb-rec:loop}). Given a specific candidate $(p,v) \in C$ the following steps are executed. First, the assignment is executed in \cref{line:bnb-rec:assign}. Given the updated assignment $\sigma$, we get a new set of candidates by the blackbox function \textsc{ExtensionCandidates()} and call the \textsc{extendSolution()} function recursively. If we found a proper solution $\sigma^\prime$ in the so created branch, we pass it down to the root of the recursion tree. Otherwise, the given assignment $\sigma$ could not be extended with the current candidate $(p,v)$ to a proper solution. Therefore, we can revert the assignment of $(p,v)$ and move on to the next candidate in $C$.

In the following sections, we describe how this Branch-\&-Bound framework, which we introduced in \Cref{alg:bnb-start,alg:bnb-rec}, can be implemented for general patterns (\Cref{sssec:proto-bnb}) and how we improved on that given specific pattern graphs $P$ like matchings (\Cref{sssec:bnb-matchings}), cycles (\Cref{sssec:bnb-cycles}), and paths (\Cref{sssec:bnb-paths}). In each of these subsections, we discuss the specific realizations of the blackbox functions \textsc{InitialCandidates}$(G)$ from \Cref{alg:bnb-start} and \textsc{ExtensionCandidates}($\sigma$, $G$) from \Cref{alg:bnb-rec}.

\subsubsection{Proto BnB}\label{sssec:proto-bnb}


We first describe a natural instantiation of our general Branch-\&-Bound-approach that can be used for any pattern graph $H$ and will be referred to as \textsc{ProtoBnb}.

\paragraph{Initial Candidates}
We wish to exploit the low-degree-dominator start to choose the first assigned vertex. Since in principle, for any low-degree-dominator $v\in N[v_{\min}]$, $v$ might be used as any pattern vertex $u\in V(H)$, a natural and safe choice is to use as initial candidates $V(H)\times N[v_{\min}]$. However, for highly symmetric pattern graphs $H$, this choice is unnecessarily wasteful, as some vertices are equivalent to each other. Thus, a better choice is to use $V'\times N[v_{\min}]$, where $V'$ is a set of a single vertex for each orbit of $H$.

For the definition of orbits, consider an arbitrary graph $H=(V,E)$. An \emph{automorphism} of $H$ is a permutation of the vertex labels such that the set of edges stays the same. Now, two vertices $v_1,v_2 \in V$ are called \emph{equivalent} in $H$ if there is an automorphism mapping $v_1$ to $v_2$. The equivalence classes of this relation are called \emph{orbits}. Now, for a given pattern $H$, we calculate the orbits of $H$. For each orbit, we add exactly one representative to the set $V^\prime$ of candidates for the pattern vertices. This is safe, because we are covering all vertices of $H$ or at least one equivalent vertex. 

To compute the orbits of a given pattern, we use the respective routine in the graph-isomorphism program \textsc{Nauty} by McKay and Piperno \cite{McKayNauty14}. The time necessary for this preprocessing step is negligible for smaller patterns, which we focus on in this paper.

\paragraph{Extension candidates}
Let $(u_0,v_0)\in V(H)\times  V(G)$ be the first assigned vertex-pair. Fix an arbitrary ordering $\pi: V(P) \to [|V(P)|]$ of the pattern vertices with $\pi(u_0) = 1$. We will successively extend our partial assignment by choosing vertices for each $u\in V(P)$ in ascending order of $\pi(u)$.

Specifically, at recursion depth $i \in \{2,\dots,|V(P)|\}$ our extension candidate set will consist of vertex pairs $\{\pi^{-1}(i)\} \times S$ for some subset $S\subseteq V(G)$ that can be assigned to the pattern vertex $\pi^{-1}(i)$. For the formal definition of the branching rule, let $\ell \in [|V(P)|-1]$ be the current recursion depth and let $p_1,\dots,p_{\ell-1}$ with $p_i \coloneqq \pi^{-1}(i)$ (for $i \in [\ell-1]$) be the pattern vertices that already have an assignment in the current partial solution $\sigma$. Then, the candidates for pattern vertex $p_\ell \coloneqq \pi^{-1}(\ell)$ are:
\begin{align}
    \left(\bigcap_{p \in N_{<\ell}} N_G(\sigma(p))\right) \;\setminus\; \left(\bigcup_{p \in \overline{N_{<\ell}}} N_G(\sigma(p))\right)\label{eq:guesses}
\end{align}
where $N_{<\ell} \coloneqq \{p \in V(G) \mid \pi(v_p) < \ell, v_p \in N_P(p_\ell)\}$ and $\overline{N_{<\ell}} \coloneqq \{p \in V(G) \mid \pi(v_p) < \ell, v_p \notin N_P(p_\ell)\}$ are the pattern vertices with assignments that are neighbors of $p_\ell$ and non-neighbors of $p_\ell$ respectively. Note that it is possible for these candidates to contain already assigned target vertices, which we  exclude separately.

\paragraph{Implementation Details}
We implemented the \textsc{ProtoBnb}-solver in a recursive manner. The crucial part here is to calculate the next set of candidates as fast as possible. For that, we store the target graph as an adjacency list where every vertex has a neighbor-array that is sorted by vertex-id. To compute the candidate for a given partial solution $\sigma$, we only need to intersect such sorted neighbor-arrays or determine the difference of two of those arrays. Both operations are possible in linear time of the size of the two given arrays. In total, we need $\ell-1$ such operations to get the candidate list for pattern vertex $p_\ell = \pi^{-1}(\ell)$. 

In order to speed up the calculation, we introduce a cache that stores common intermediate results for the candidates of the same recursive call. If $G(\ell,v_{\ell-1})$ is the set of candidates from \Cref{eq:guesses} for pattern vertex $p_\ell$ and if the target vertex $v_{\ell-1}$ was assigned to the previous pattern vertex $p_{\ell-1}$, then most of the calculations can be reused. More precisely, the result of all intersections and differences up to the pattern vertex $p_{\ell-2}$ can be reused for the next candidate $v^\prime_{\ell-1}$ for $p_{\ell-1}$ (if there is one). Therefore, we save this intermediate candidate list for every recursion depth once it was calculated during the run of the first candidate of this call and reuse it for the upcoming candidates in the same call.

The candidate list in \Cref{eq:guesses} ensures that at recursion depth $|V(H)|$, the current partial assignment actually forms the desired pattern $H$ in the host graph. To check if this complete assignment is also dominating, we use a precalculated adjacency matrix (for closed neighborhoods) and check if the OR-ed rows of the assigned target vertices contain any $0$ entry.


\subsubsection{BnB-Solver for Matchings}\label{sssec:bnb-matchings}
In this section, we improve our Branch-\&-Bound solver for dominating $k$-matchings, i.e., matchings with $k$ edges (for a fixed $k$). 

The main idea is to integrate the domination aspect of the problem as frequently as possible, by including it appropriately into our branching rules. Specifically, as observed in Section~\ref{ssec:heuristics}, whenever a partial solution leaves at least one node not dominated, we can pick the undominated node $v$ of minimum degree and branch over $v'\in N[v]$ for inclusion into our partial solution. (We will do so only if this results in fewer branches than extending the pattern as in \textsc{ProtoBnb}.) However, in contrast to \textsc{ProtoBnb}, we lose control over which pattern node will be assigned to $v'$. One option would be to choose as candidate set $V_\bot\times N[v]$, where $V_\bot\subseteq V(H)$ denotes the subset of \emph{unassigned} pattern vertices. However, we notice that for patterns such as matching, we can improve over this approach by determining the \emph{role} of $v$ on the spot. In the remainder of this section, we describe the details of this approach for matchings.

Note that in the beginning, all pattern vertices are in the same orbit (are equivalent). Therefore, we can use $\{p\}\times N[v_{\min}]$ for any pattern vertex $p$ as our initial set of candidates. 
In general, we interpret every partial assignment as a set of pattern edges $e=(u,v) \in E(H)$ that are either \emph{not yet started}, \emph{started} or \emph{finished} depending on whether none, one or both of $u$ and $v$ already have been assigned to a target vertex in the partial solution. During the algorithm we always keep track of the started pattern edges as well as the number of yet unstarted edges. Based on a given partial assignment $\sigma$, we use two different branching rules to extend $\sigma$.
\begin{enumerate}
    \item \textsc{DominationRule}: We know that every target vertex has to be dominated in the end. Therefore, we can branch over the inclusive neighborhood $N_G[v]$ for some target vertex $v$ that is not dominated by any vertex in $\sigma$. To reduce the branching factor, we choose the undominated vertex with the smallest degree.
    \item \textsc{ExtensionRule}: Since we are looking for a matching in the target graph, we know that every started edge in $\sigma$ has to be finished in the end. Therefore, we can branch over the (open) neighborhood $N_G(v)$ of every target vertex $v$ that is part of a started edge in $\sigma$. Again, to reduce the branching factor, we choose the one with the smallest degree.
\end{enumerate}

\begin{figure}
    \centering
    \includegraphics[width=0.6\linewidth,page=1]{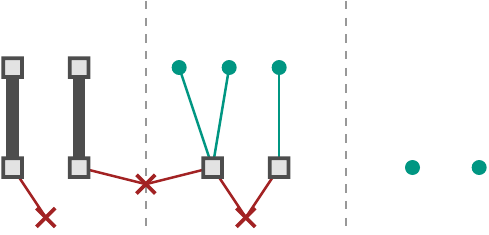}
    \caption{Sketch for possible ways to extend a partial assignment for $k$-matching pattern. The black squares are vertices of the host graph that are assigned to some pattern vertex. The other marks show how this partial assignment can be extended properly (green dots) or incorrectly (red crosses). Among all these vertices all those edges (of the host graph) are shown that are incident to some assigned vertex.}
    \label{fig:matching-sketch}
\end{figure}

\noindent
For \textsc{ExtensionCandidates()} we choose the smallest candidates list of these two branching rules. For every target vertex $v$ in that candidate list, we determine in which way it would extend the current partial assignment. For that, let $D$ be the number of assigned target vertices in $\sigma$ that dominate $v$. If $D=0$, we start a new pattern edge (if possible). If $D=1$, we know that $v$ finishes exactly one started pattern edge. For every $D\geq 2$, there is no valid way for $v$ to extend $\sigma$, since every vertex in a matching has degree of at most $1$.

\paragraph{Common Implementation Details.}
We implemented the Branch-\&-Bound solver for matchings as well the ones for cycles and paths (described below) in an iterative manner by using an explicit call stack. This explicit stack has exactly size $\ell$ when $\ell-1$ pattern vertices are already assigned and the $\ell$\,th pattern vertex is about to get its assignment. Every entry of the stack stores a list of the candidates that are iterated in the respective recursion depth. 

All of the pattern-specific BnB-solvers use the \textsc{DominationRule} (introduced in \Cref{sssec:bnb-matchings} about the matching solver) where the (inclusive) neighbors of a yet undominated target vertex are the potential candidates to extend the pattern. To find the undominated vertex with the smallest degree fast, we sort the target vertices increasingly by degree in a preprocessing step and do a linear search on the sorted list. Again, we can use a cache for every recursion depth storing the index of the smallest undominated vertex, since these indices are only increasing with increasing recursion depth.

To figure out how a given target vertex $v$ can extend the current partial solution, we keep track of the number of assigned target vertices in $\sigma$ that dominate $v$ in $G$ (for every $v \in V(G)$). Whenever a new target vertex $u$ is assigned, this number is incremented for all its (inclusive) neighbors, and it is decremented again when $u$ is unassigned.

A slight disadvantage of only tracking the number of dominating assigned vertices for every pattern vertex $v$ is that one has no access to the list of those vertices itself. Therefore, an additional search is necessary to find the $D$ many dominating assigned vertices that $v$ is adjacent to. For example in the Matching solver: when we know that a target vertex $v$ is dominated by $D=1$ assigned target vertices, we do not know which pattern edge $v$ completes. To make this search or specifically the needed adjacency tests faster, we store for every target vertex $v$ an additional hash containing the IDs of all of $v$ neighbors in $G$. We do the same for the other pattern-specific BnB-Solver described in \Cref{sssec:bnb-cycles,sssec:bnb-paths}, as similar problems arise there, too.

\subsubsection{BnB-Solver for Cycles}\label{sssec:bnb-cycles}
In this section, we describe our tailored Branch-\&-Bound solver for $k$-cycles, i.e., cycles with $k$ vertices (and edges). As already noted in \Cref{sssec:bnb-matchings}, our BnB-Solver for cycles shares the same techniques as the matching solver including the explicit call stack, the tracking of the number of dominating vertices per vertex and the hash sets in addition to the adjacency lists.

As a main difference to the matching solver, the partial assignment for cycles is a bit more complex and needs more bookkeeping. Every partial assignment of a $k$-cycle is a set of induced paths in the target graph that we call \emph{(cycle) segments}. We store each segment as doubly-linked list of the target vertices. This way, we can easily join segments or remove a vertex in the middle of some segment while still having quick access to the ends of the segments. The end of the segments are important here, as they are the only options to extend the partial assignment properly. 

Given a partial assignment $\sigma$ with cycle segments $P_1,\dots,P_\ell$ we use two branching rules: the \textsc{DominationRule} (unchanged from \Cref{sssec:bnb-matchings}) and the following variant of the \textsc{ExtensionRule} adapted for cycles. We know that, at the end, there may only be one cycle segment $P$ in $\sigma$ whose ends have to be adjacent. Therefore, it is necessary to close all open segment ends, in order to complete the current partial assignment. As a consequence, we can safely branch over the exclusive neighborhood of every target vertex at the end of one of the cycle segments $P_1,\dots,P_\ell$. Again, to reduce the branching factor, we choose the end with the smallest degree to branch on. In the end, we chose the branching rule that yields the smallest branching factor in each iteration.

Let $v$ be a target vertex and $\sigma$ the partial solution with $P_1,\dots,P_\ell$. Then, there are four option on how $v$ can extend $\sigma$ depending on the number $D_\text{end}$ of assigned vertices in $\sigma$ that are an end of a cycle segment and dominate $v$:
\begin{itemize}
    \item If $D_\text{end}=0$, $v$ starts a new cycle segment. This is only done if the number of unassigned pattern vertices (after $v$ was assigned) is large enough to close all open ends in the now extended partial solution.
    \item If $D_\text{end}=1$, we append $v$ to $P$ either to the front or to the end. This can only be done if the number of remaining assignments is larger or equal to the number of cycle sections. Otherwise the cycle can not be closed anymore after that.
    \item If $D_\text{end}=2$, the vertex $v$ is adjacent to the ends of cycle segments $P$ and $P^\prime$. If $P$ and $P^\prime$ are different, we join them. If, on the other hand, $P$ and $P^\prime$ are the same segment, i.e., $v$ is adjacent to both ends of $P$, $v$ can only be assigned if $P$ is the only segment of $\sigma$ and $v$ is the last vertex to be assigned.
\end{itemize}

\begin{figure}
    \centering
    \includegraphics[width=0.35\linewidth,page=2]{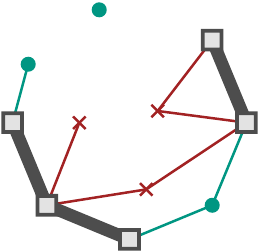}
    \caption{Sketch for possible ways to extend a partial assignment for $k$-cycle pattern. The black squares are vertices of the host graph that are assigned to some pattern vertex. The other marks show how this partial assignment can be extended properly (green dots) or incorrectly (red crosses). Among all these vertices all those edges (of the host graph) are shown that are incident to some assigned vertex.}
    \label{fig:cycle-sketch}
\end{figure}

\subsubsection{BnB-Solver for Paths}\label{sssec:bnb-paths}
For $k$-paths, i.e., paths with $k$ vertices, the BnB-approach and the implementation details stay the same as for the other pattern-specific solvers for matchings (\Cref{sssec:bnb-matchings}) and cycles (\Cref{sssec:bnb-cycles}). Additionally, a partial solution for a path is also a set of induced paths in the target graph like with the cycle solver. However, there are some important differences compared to the cycle solver. First of all, the extend option to connect two ends of the same segment (here \emph{path segments}) is not valid anymore.

Another, more important, difference is that the branching rule \textsc{ExtenionRule} as described in \Cref{sssec:bnb-cycles} is not safe for paths anymore. The reason for that is that the final structure (a path) has two ends itself. Therefore, not every open end of every path segment has to be closed. It is thus not guaranteed for some arbitrary end of a path segment that one of its neighbor has to be part of the final solution. However, for every path segment $P$ with ends $u$ and $v$, it is safe to branch over the combined neighborhoods of $u$ and $v$, i.e., $N_G(u)\cup N_G(v)$. 

For the correctness of this branching rule consider the two cases where there is only one segment and where there are multiple segments in the current partial solution $\sigma$. If there is only segment $P$ in $\sigma$ with ends $u$ and $v$ and there are still pattern vertices left to assign, then at least of the ends $u$ and $v$ is not an end of the final path (otherwise this final path based on $\sigma$ would not be an induced path). On the other hand, if there are multiple path segments $P_1,\dots,P_\ell$ in $\sigma$, then every final solution $\sigma^*$ based on $\sigma$ joins these segments together. For all segments to be joined together, every segment has at least one end that is closed in $\sigma^*$. As a consequence, it is safe to branch over the combined neighborhood of the ends of one path segment. Again, we choose the segment with the smallest combined neighborhood (of its ends) and choose the branching rule that yields the smallest set of candidates.




\section{Experimental Evaluation}
In this section, we compare the runtime of our Branch-\&-Bound solvers (from \Cref{ssec:bnb}) to the other baseline approaches (described in \Cref{ssec:baseline-approaches}). In \Cref{ssec:eval-highlevel}, we evaluate all solvers on different benchmarks for small matchings, cycles and paths as pattern. In \Cref{ssec:eval-detailed}, we compare the runtimes of SAT, eGSS, and our solver for matchings in more detail, focusing primarily on the influence of pattern size on runtime.

\subsection{Evaluated Datasets}

As real-world benchmarks for the Dominating $H$-Pattern problem, we create benchmark sets by selecting non-trivial instances from various sources. Generally, we aim for most instances being reasonable candidates for containing a dominating $H$-pattern. To achieve this, we usually select graphs which have domination number at most $|V(H)|$ (or only slightly larger) and usually contain at least one occurrence of an induced $H$ -- this way, we focus on instances where existence of a dominating $H$-Pattern is not already precluded by simple reasons.
All of our graphs are undirected. Our benchmarks are:

\begin{itemize}
    \item \textsc{rome}: a collection of 11,534 graphs with 10-110 nodes (69.1 edges on average) taken from a graph drawing dataset~\cite{rome-graphs}.
    \item \textsc{HoG-M3}: the first 2000 graphs returned by House of Graphs (HoG)~\cite{house-of-graphs} on querying graphs having domination number at most $6$, at least 20 vertices and containing an induced $M_3$.
    \item HoG-M3-5-hard: the first 1494 graphs returned by HoG on querying graphs having domination number at most $6$, a minimum degree at least 5 and containing an induced $M_5$ (timeout of 60s for the pattern detection).
    \item HoG-ds1-10-hard: all 1401 graphs returned by HoG on querying graphs having domination number at most $10$, at least 20 vertices and minimum degree at least $6$
    \item \textsc{stride}: a collection of 250 graphs returned by the Stride dataset~\cite{stride} on querying graphs with dominating set size at most 6, at least 100 vertices and at most 100,000 edges. 
\end{itemize}

\paragraph{Synthetic hard instances}
We also generate hard instances using fine-grained reductions, resulting in a benchmark set modeling worst-case behaviors (\textsc{hard OV}).
Specifically, for each investigated pattern $H$, we generate a collection of 100 randomly generated $|V(H)|$-Orthogonal Vectors instances that are then reduced to the Dominating $H$-Pattern problem using the reduction in \cite{DransfeldKR25}.
Each entry in every vector has an independent 75\% chance of being set to $1$.
For NO-instances, dimensions are added until the OV instance does not have a solution anymore.
For YES-instances, the procedure for NO-instances is used, and the last dimension is then deleted.
This procedure is designed to create few vertices of low degree after the reduction.
The sizes of the vector sets were chosen depending on the pattern to create hard instances for the fastest solvers.

\subsection{Experimental Setup}
We ran all our experiments on a Supermicro SuperServer SYS-6029UZ-TR4+ server with an Intel(R) Xeon(R) Gold 6144 CPU @ 3.50GHz processor and 192GB DDR4 (2666MHz) memory. All runs had a timeout of five minutes.

The implementation of all solvers as well as all benchmarks and result files are available on Zenodo\footnote{10.5281/zenodo.17344937}. 
Our Branch-\&-Bound solvers are written in C++17 and compiled with version 14.2.0 of gcc with the -O3 flag enabled. The eGSS and the kissat solver are written in C++ and C respectively, and were compiled with the same settings. We used the following versions for the different baseline solvers:
\begin{itemize}
    \item Glasgow-subgraph-solver (GSS): commit \texttt{\#68ea3d1}
    \item kissat (SAT): version 4.0.3
    \item Gurobi (ILP): version 12.0.2
\end{itemize}


	


\subsection{High-level comparison}\label{ssec:eval-highlevel}
In this subsection, we evaluate the runtime of all solver on benchmarks for small matchings, cycles and paths of almost equal size. Specifically, we consider $M_3$, $C_5$ and $P_5$. We compare the performance of our \textsc{ProtoBnb} solver and our tailored BnB-solver (for the respective pattern) to the baseline approaches of SAT (with default \amoc-encoding), ILP and eGSS. 

\begin{figure*}[ht]
\begin{subfigure}[t]{0.5\textwidth}
\includegraphics[width=\textwidth]{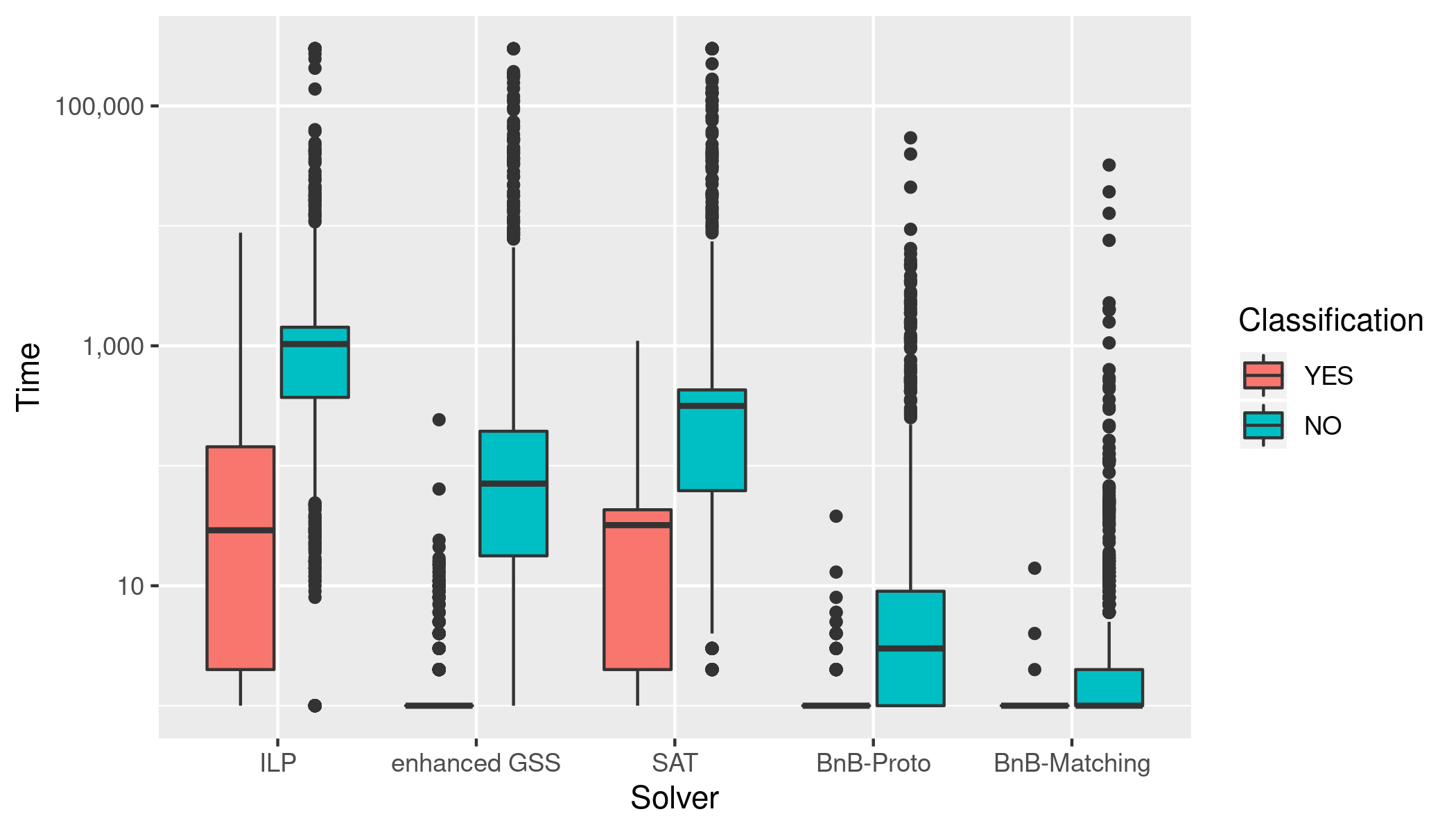}
\caption{HoG-ds1-10-hard}
\end{subfigure}
\begin{subfigure}[t]{0.5\textwidth}
\includegraphics[width=\textwidth]{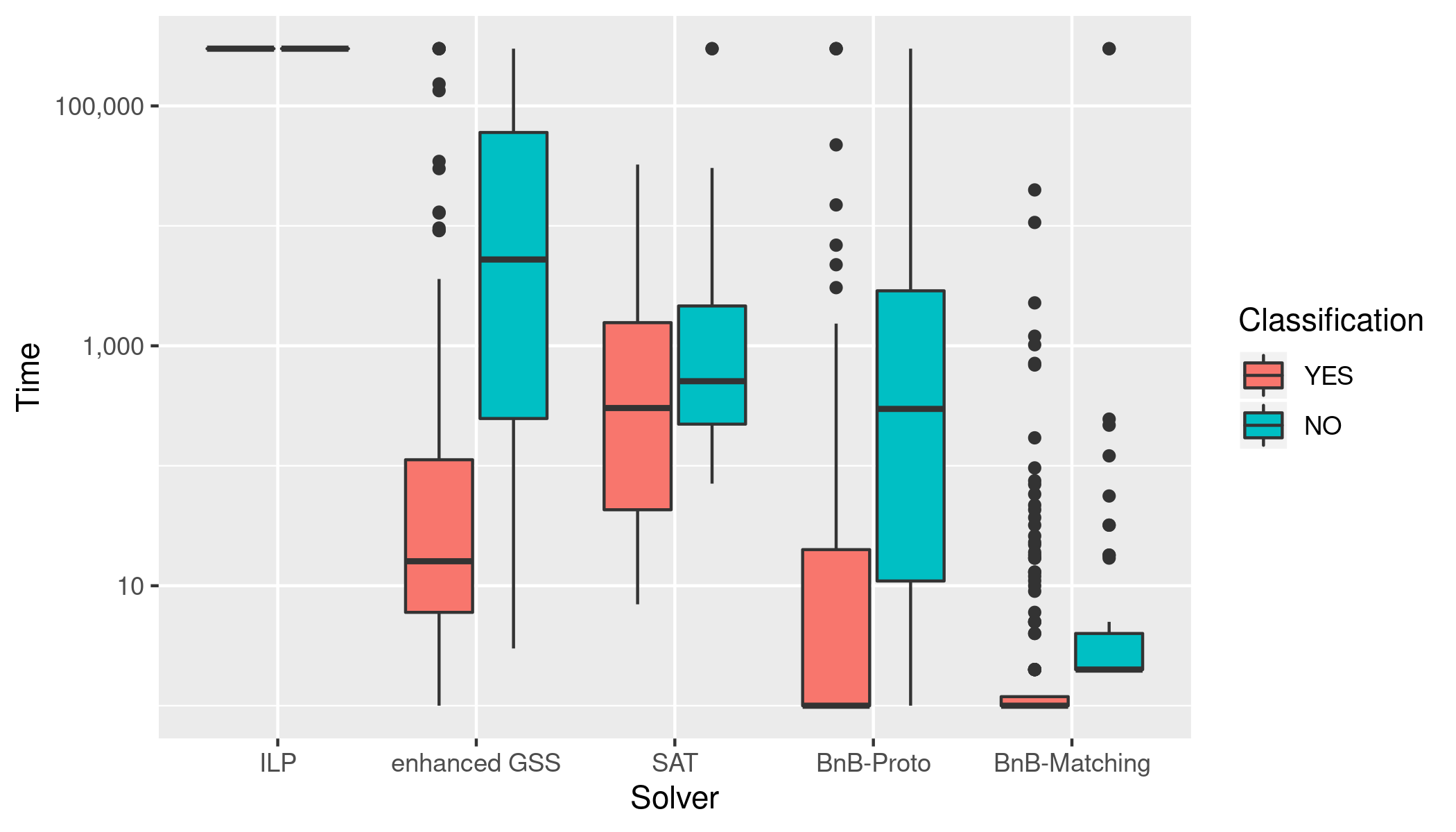}
\caption{stride}
\end{subfigure}
\caption{Running times for Dominating $M_3$-Pattern on different datasets in ms.\label{fig:M3}}
\end{figure*}

\paragraph{Matchings} 
The full results for the $M_3$-pattern are given in the appendix in \Cref{fig:appendix-M3} and in~\Cref{tab:M3}. \Cref{fig:M3} shows the runtimes of all solver for the stride and the HoG-ds1-10-hard benchmarks. 

On all benchmarks of small real-world graphs (Rome and HoG-M3) both our BnB-solvers solve all instances in under 10ms, and thus outperform SAT and ILP by at least one order of magnitude on these instances. This still holds true for the matching-BnB solver on larger real-world graphs (stride) an the more complex HoG graphs. The \textsc{ProtoBnb} solver, however, becomes slower for larger and more complex graphs. On the OV-instances, the matching-BnB solver even beats the \textsc{ProtoBnb} by at least two orders of magnitude.

The OV-instances are shown to be difficult to solve, since all the baseline solvers run into timeouts for NO-instances and sometimes even on YES-instances. Among the baseline solvers, the eGSS performs best on small real-world instances (Rome and HoG-M3) and on the more complex instances from the HoG. On OV-instances, however, it is significantly slower than the SAT-solver and (at least on the YES-instances) even slower than the ILP-solver. On the stride instances, it is unclear which of the eGSS and SAT-solver to prefer, since eGSS solve YES-instances faster than SAT, but SAT slightly wins on the NO-instances.

Our tests with the ladder encoding for SAT revealed that perhaps surprisingly, the smaller encoding led to generally slower solve times so that we focus here purely on the default encoding for SAT.

To conclude the $M_3$ results, our solvers are always among the fastest while the tailored BnB-solver is consistently faster than the \textsc{ProtoBnb}. Enhanced eGSS achieves similar runtimes on the smaller real-world graphs but loses significantly on larger (stride) and more complex graphs, where the SAT-solver performs best among the baseline approaches. Over all benchmarks and independent of whether we look at YES- or NO-instances, the ILP-solver is the slowest solver.

\begin{figure*}[ht]
\begin{subfigure}[t]{0.5\textwidth}
\includegraphics[width=\textwidth]{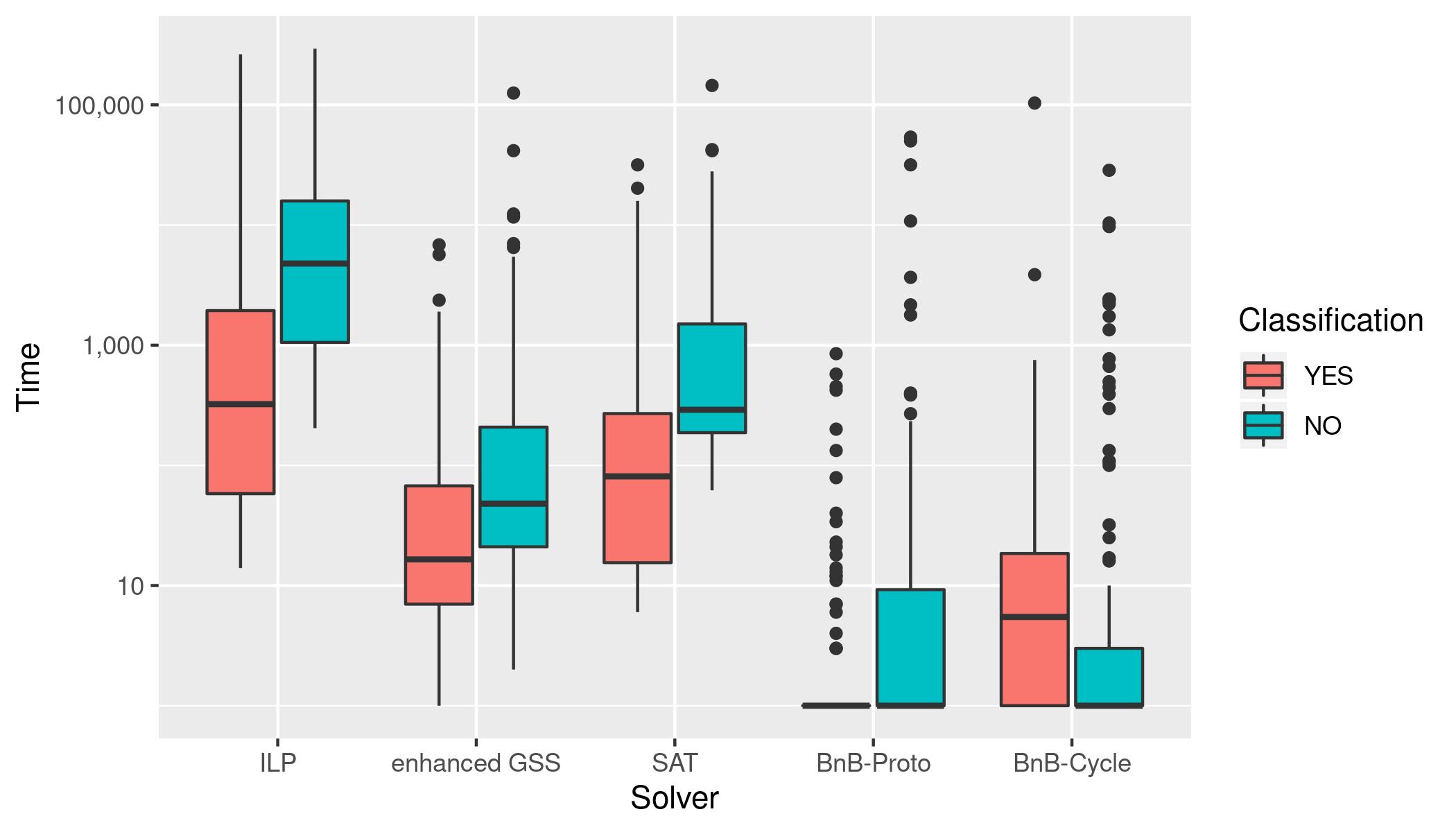}
\caption{stride, $C_5$}
\end{subfigure}
\begin{subfigure}[t]{0.5\textwidth}
\includegraphics[width=\textwidth]{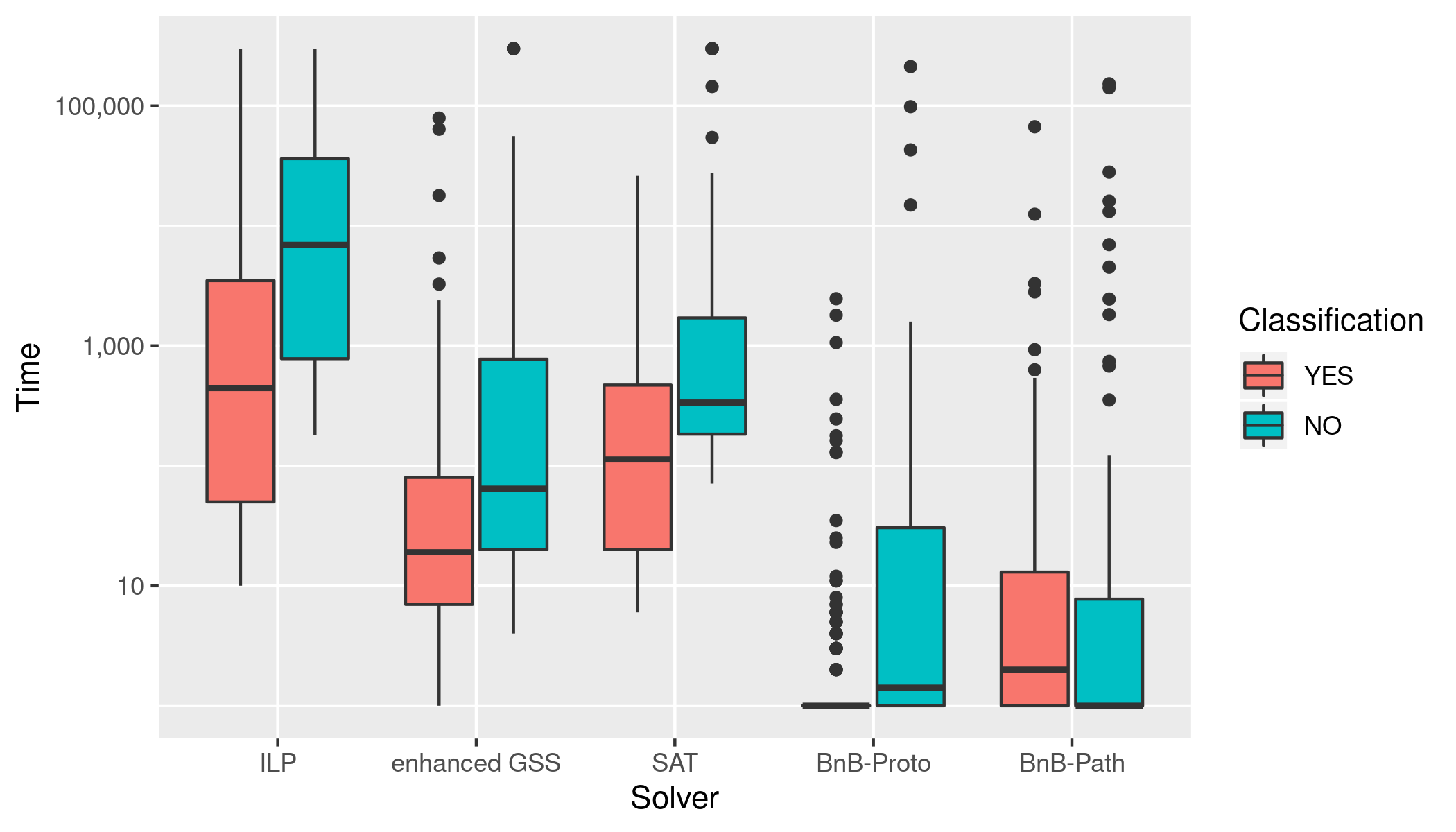}
\caption{stride, $P_5$}
\end{subfigure}
\caption{Running times for Dominating $C_5$-Pattern (left) and $P_5$-Pattern on the stride dataset in ms.\label{fig:C5-P5}}
\end{figure*}

\paragraph{Cycles}
The results for the $C_5$-pattern are given in the appendix in \Cref{fig:appendix-C5} and in~\Cref{tab:C5}. Here, the solvers behave nearly consistently among all test benchmarks. As an example, \Cref{fig:C5-P5} (a) shows the results for the stride benchmark. The results show that eGSS performs better than SAT on realistic benchmarks and that both solvers beat the ILP as the worst performing solver. The largest difference can be seen on the Rome NO-Instances, where the eGSS solve almost all instances instantly ($<$1ms), while the other baseline solvers have a median runtime of about 50ms (SAT) and 200ms (ILP). Here, both of our solvers perform just as well as the eGSS. As with the $M_3$-pattern, both of our BnB-solver are either among the best solver are the only top-performing solver over all benchmarks. Interestingly, the \textsc{ProtoBnb} is slightly faster than the BnB-solver that was explicitly tailored for the cycle pattern. Here, the YES-instances of the larger real-world graphs (stride) yield the highest discrepancy.

\paragraph{Paths}
On the $P_5$-pattern, the results are nearly identical to the ones of the $C_5$-pattern with only minor changes in the actual runtimes.
The full results for the $P_5$-pattern are given in the appendix in \Cref{fig:appendix-P5} and in~\Cref{tab:P5}, and an example is shown 
in \Cref{fig:C5-P5} (b).

\paragraph{Conclusion}
Over all tested patterns, at least one of our Branch-\&-Bound solvers is among the fastest. For $M_3$, our tailored $M_3$-BnB-solver clearly beats the other solvers. On $C_5$ and $P_5$, however, the \textsc{ProtoBnb} has a slight advantage over the tailored versions. Among the other baseline approaches, the ranking of the solvers is clear for almost all benchmarks: eGSS is faster than SAT, which is faster than ILP. The only exception to this are the larger and more complex instances for the $M_3$-pattern, where the SAT-solver is either equivalently fast (stride) or even beats the eGSS by at least one order of magnitude (YES-instances of OV-benchmark).










\subsection{Influence of Pattern}\label{ssec:eval-detailed}

\begin{figure*}[t]
\begin{subfigure}[t]{0.5\textwidth}
\includegraphics[width=\textwidth]{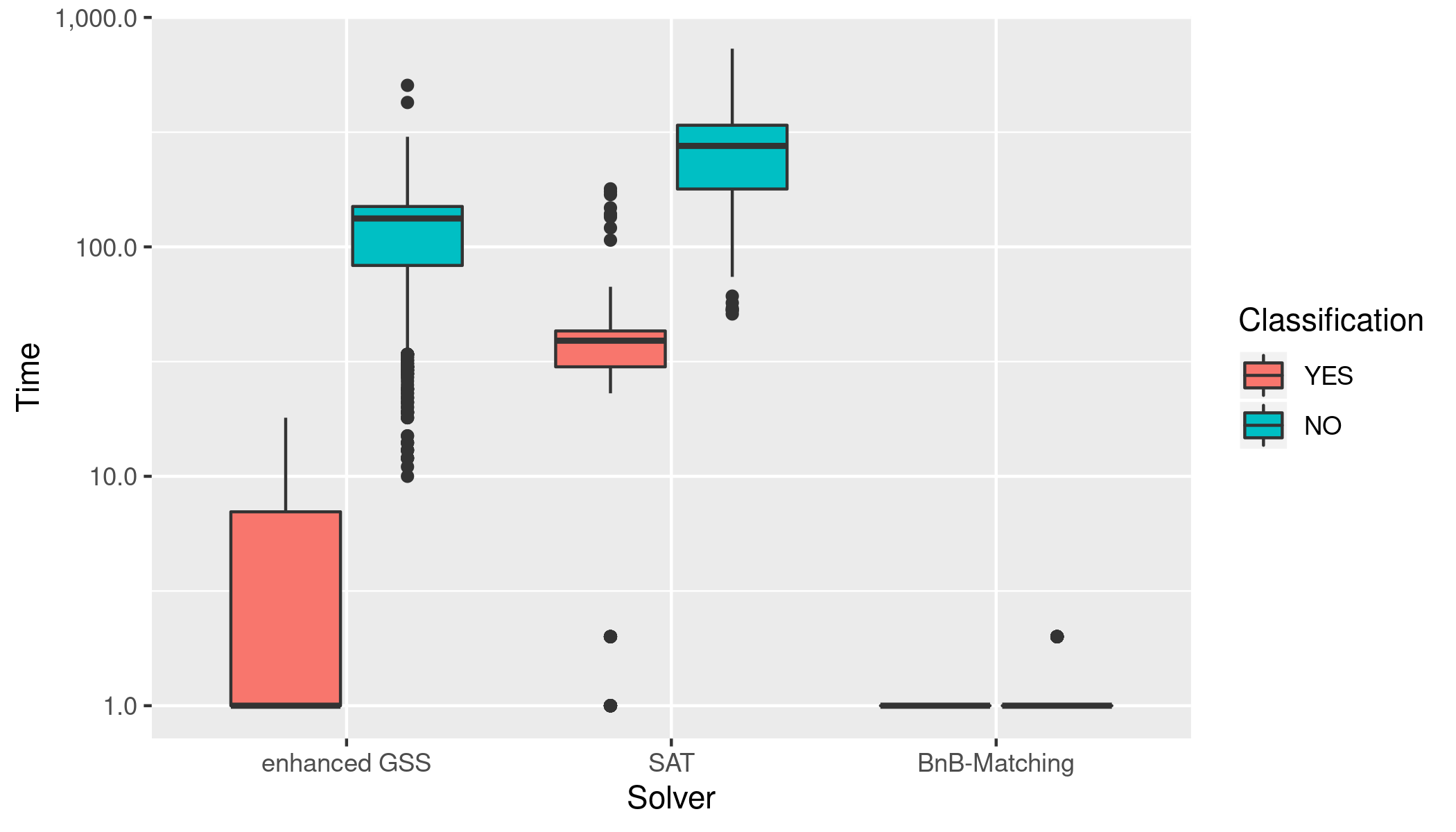}
\caption{$M_3$ on HoG-M3-5-hard}
\end{subfigure}
\begin{subfigure}[t]{0.5\textwidth}
\includegraphics[width=\textwidth]{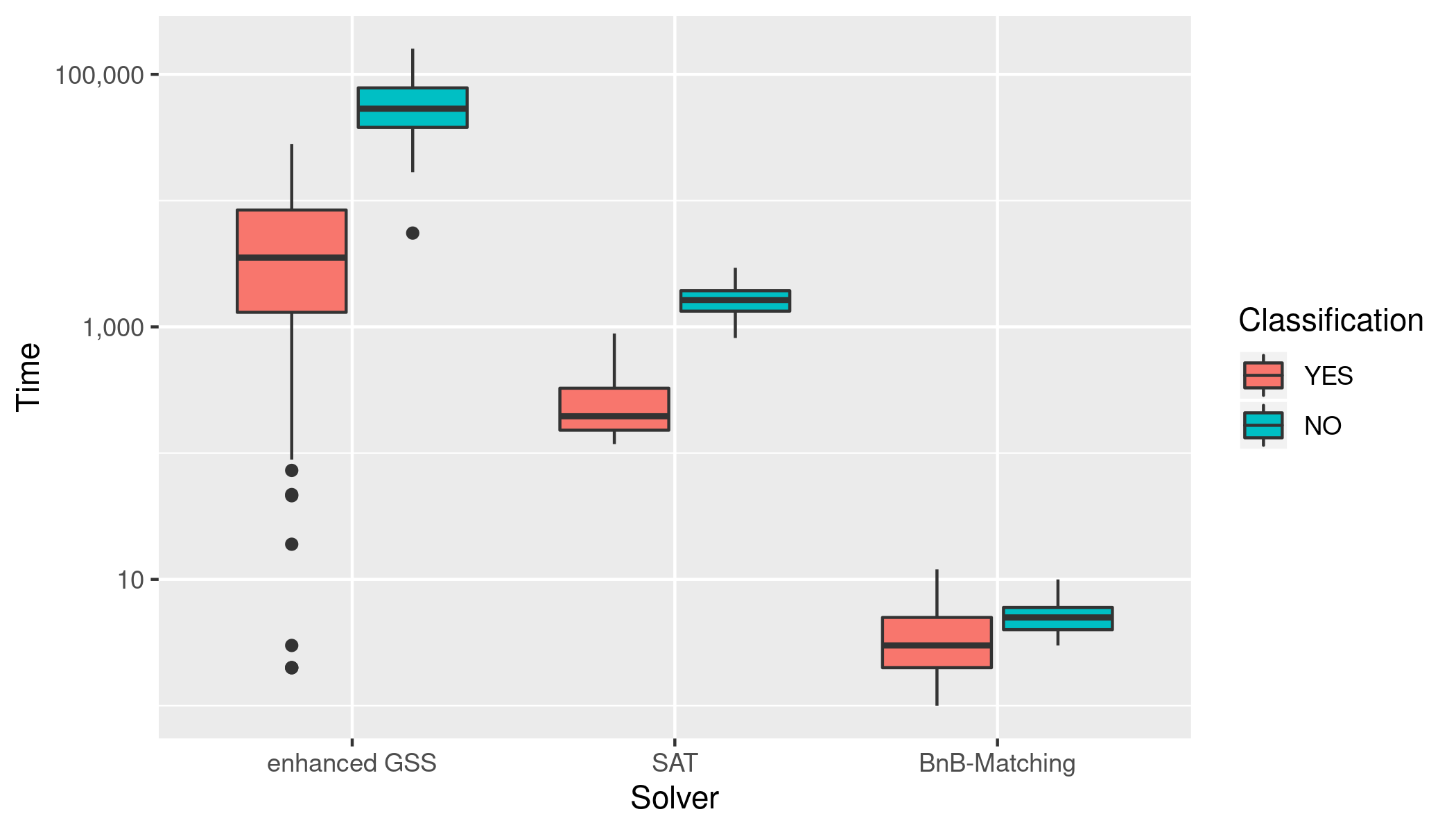}
\caption{$M_3$ on OV instances}
\end{subfigure}

\begin{subfigure}[t]{0.5\textwidth}
\includegraphics[width=\textwidth]{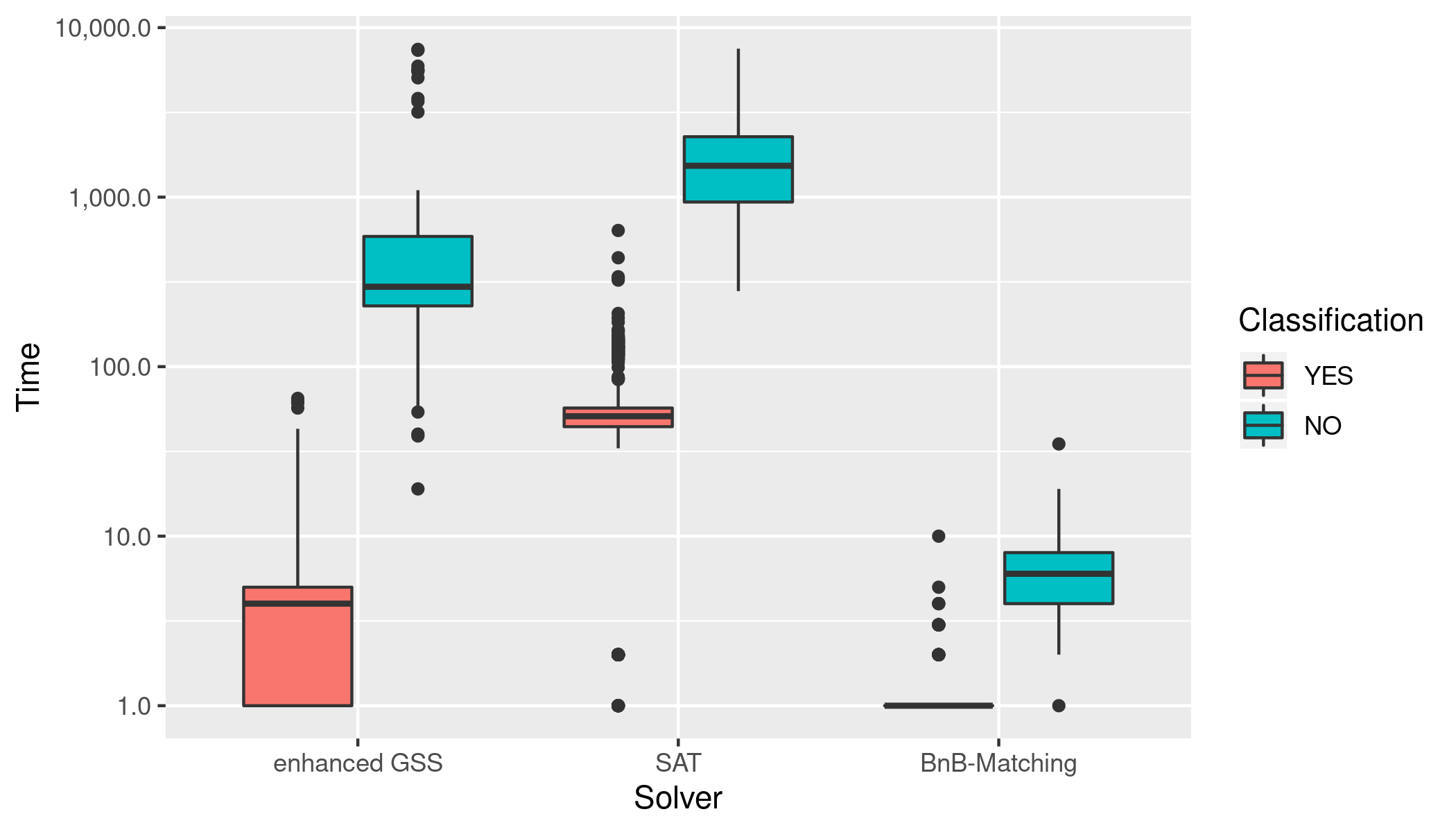}
\caption{$M_4$ on HoG-M3-5-hard}
\end{subfigure}
\begin{subfigure}[t]{0.5\textwidth}
\includegraphics[width=\textwidth]{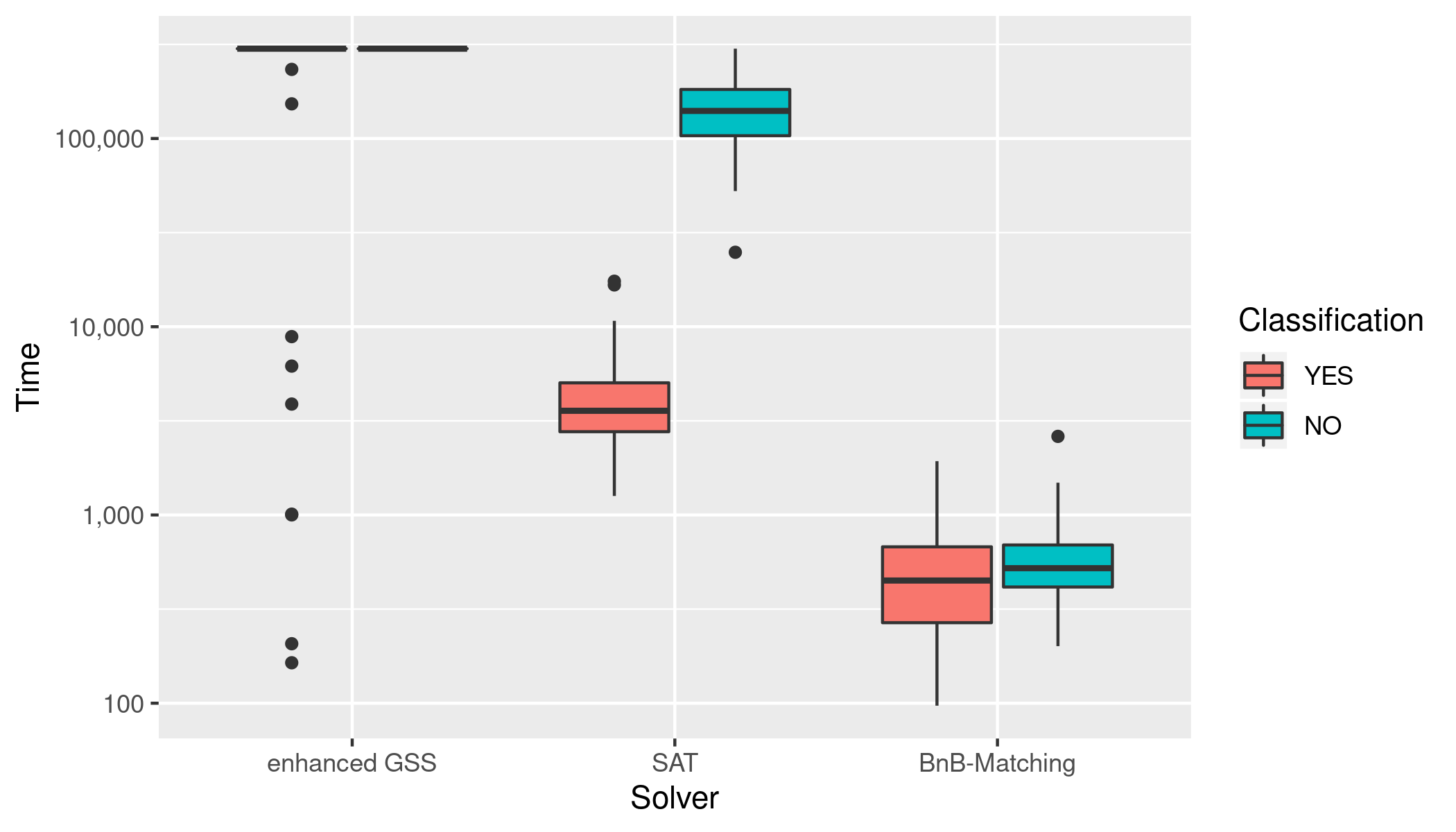}
\caption{$M_4$ on OV instances}
\end{subfigure}

\begin{subfigure}[t]{0.5\textwidth}
\includegraphics[width=\textwidth]{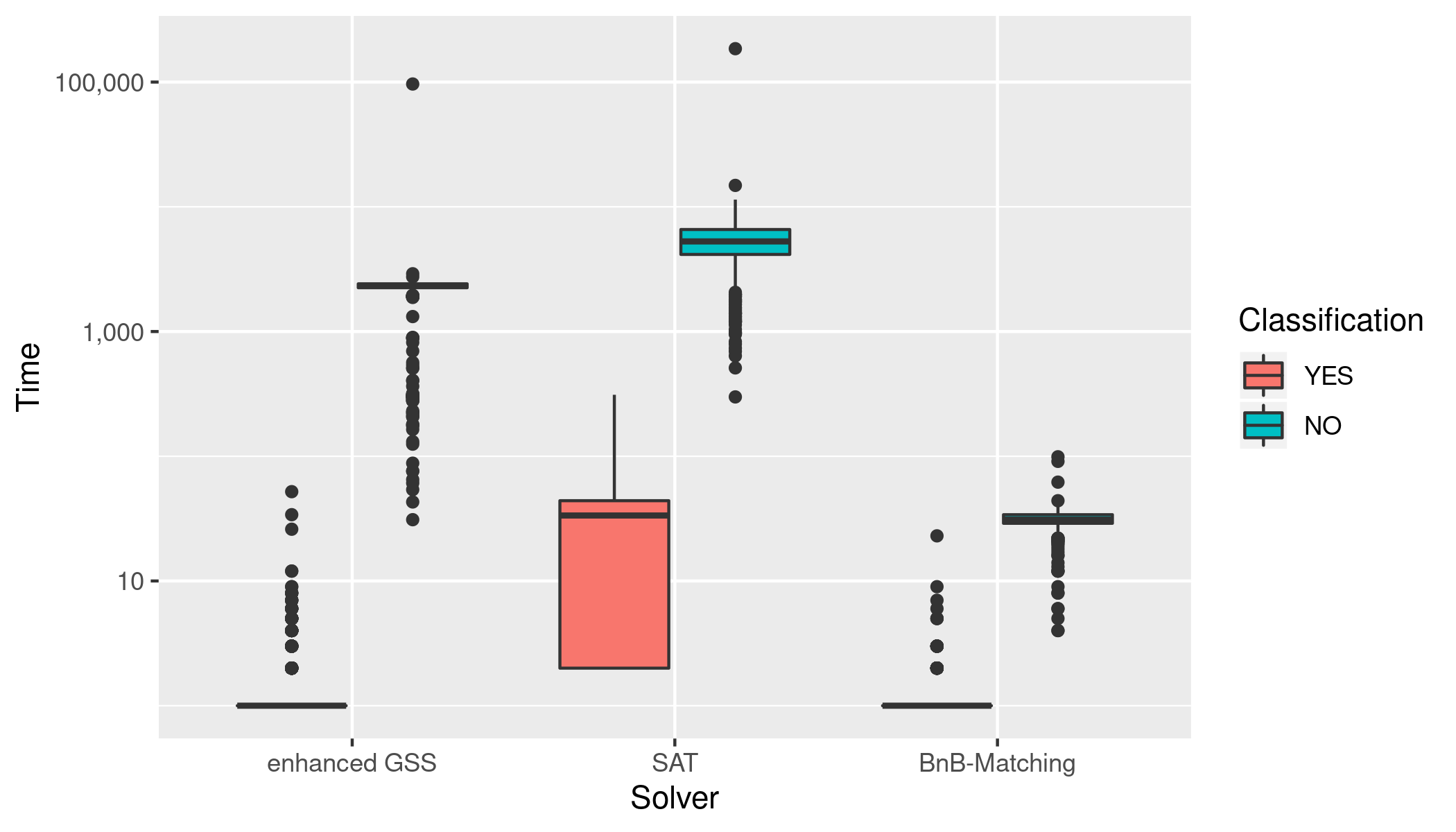}
\caption{$M_5$ on HoG-M3-5-hard}
\end{subfigure}
\begin{subfigure}[t]{0.5\textwidth}
\includegraphics[width=\textwidth]{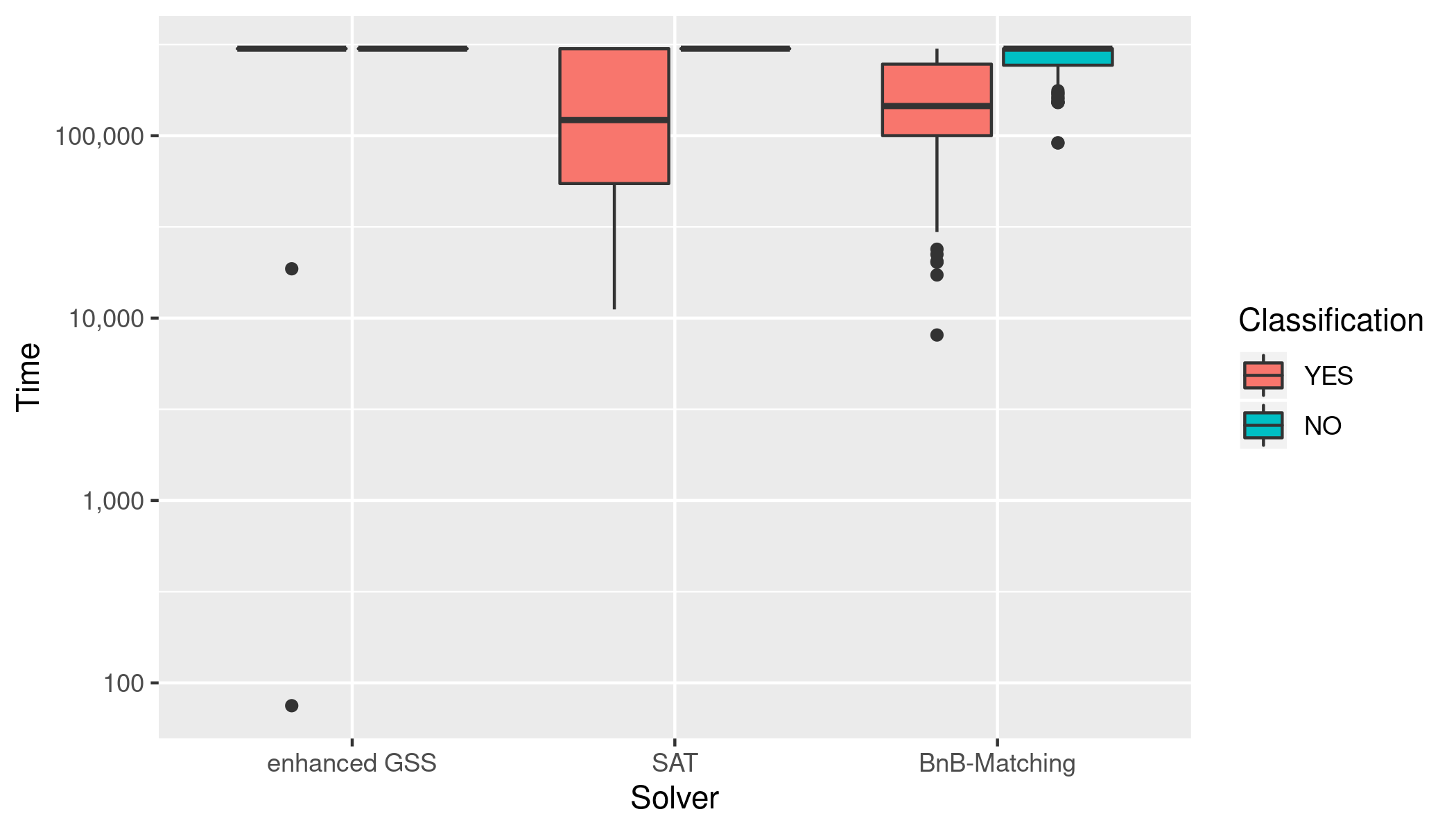}
\caption{$M_5$ on OV instances}
\end{subfigure}

\caption{Influence of increasing the pattern size for matchings $M_3, M_4, M_5$. The left column is on the same dataset HoG-M3-5-hard. The generated OV instances in the right column increase in size with matching size.\label{fig:M-progression}}
\end{figure*}


In the appendix in Table~\ref{fig:M-progression}, we provide additional plots that highlight the dependence of the running times on the pattern size. Here, we give data for matchings $M_3, M_4, M_5$ on (1) the fixed real-world dataset HoG-M3-5-hard (which contains reasonable instances for each such matching size),  (2) generated OV instances, where the instances are created adaptively to the pattern, i.e., increase in size with the pattern size. We observe that our Branch-\&-Bound approach consistently outperforms enhanced eGSS and kissat, and the running times increase indeed substantially with the matching size. Indeed, on the NO-instances of the dataset HoG-M3-5-hard, the median runtimes of SAT and eGSS increase by at least factor of 3 and 7.8 respectively when comparing the $M_5$- to $M_4$-results.











\section{Final Remarks}
In this paper we describe the first practical implementation of Dominating $H$-Pattern problem which generalizes two classical problems from graph theory: (1) Dominating Set, (2) Pattern Detection.
Our main algorithmic contribution is constructing a Branch-\&-Bound algorithm that combines these two aspects of the problem, and by using a careful interplay between the two aspects, we are able to significantly improve the running times when compared to the other solvers from the literature.
Our experimental results suggest that, among the generic solvers, the SAT solver is a more robust choice.
However, on many real-world instances, it is still outperformed by the enhanced Glasgow Subgraph Solver (eGSS).
Notably, our Branch-\&-Bound algorithms consistently achieve significant performance improvements over all other investigated approaches on both realistic and synthetic hard benchmarks.

Future research includes the development of a specialized solver capable of efficiently handling arbitrary pattern graphs. 

\bibliographystyle{plainurl}
\bibliography{refs}

\appendix
\section{Experimental Data}
    \clearpage
    \thispagestyle{empty}
    \begin{landscape}

        \centering 

\begin{table}
\footnotesize
\begin{tabular}{l|rr@{$\;$}r@{}rr@{$\;$}r@{}r|r@{$\;$}r@{}r|r@{$\;$}r@{}rr@{$\;$}r@{}r|r@{$\;$}r@{}rr@{$\;$}r@{}r}
        Dataset & Size & \multicolumn{3}{c}{$n$} & \multicolumn{3}{c}{eGSS} & \multicolumn{3}{c}{kissat}  & \multicolumn{3}{c}{Gurobi} & \multicolumn{3}{c}{proto-BnB} & \multicolumn{3}{c}{BnB-Matching}\\
        \hline
rome-y & 538 & 13 & ([10, & 20]) & 0 & ([0, & 0]) & 0.001 & ([0.001, & 0.029]) & 0 & ([0, & 0.213]) & 0 & ([0, & 0]) & 0 & ([0, & 0])\\
rome-n & 10994 & 49 & ([10, & 110]) & 0 & ([0, & 0.016]) & 0.049 & ([0.002, & 0.127]) & 0.202 & ([0, & 5.78]) & 0 & ([0, & 0]) & 0 & ([0, & 0])\\
HoG-M3-y & 401 & 20 & ([20, & 23]) & 0 & ([0, & 0.002]) & 0.023 & ([0.001, & 0.041]) & 0.022 & ([0, & 0.5]) & 0 & ([0, & 0]) & 0 & ([0, & 0])\\
HoG-M3-n & 1597 & 22 & ([20, & 23]) & 0.018 & ([0, & 0.057]) & 0.031 & ([0.022, & 0.247]) & 1.511 & ([0, & 13.043]) & 0.001 & ([0, & 0.002]) & 0 & ([0, & 0.001])\\
HoG-ds1-10-hard-y & 471 & 28 & ([20, & 171]) & 0 & ([0, & 0.242]) & 0.032 & ([0.001, & 1.102]) & 0.029 & ([0, & 8.783]) & 0 & ([0, & 0.038]) & 0 & ([0, & 0.014])\\
HoG-ds1-10-hard-n & 928 & 28 & ([20, & 162]) & 0.071 & ([0.001, & TO]) & 0.316 & ([0.002, & TO]) & 1.035 & ([0, & TO]) & 0.003 & ([0, & 54.198]) & 0.001 & ([0, & 32.147])\\
stride-package1-y & 168 & 184 & ([100, & 1329]) & 0.016 & ([0, & TO]) & 0.303 & ([0.007, & 32.44]) & TO & ([TO, & TO]) & 0.001 & ([0, & TO]) & 0 & ([0, & 19.937])\\
stride-package1-n & 79 & 180 & ([102, & 964]) & 5.244 & ([0.003, & TO]) & 0.507 & ([0.071, & TO]) & TO & ([TO, & TO]) & 0.298 & ([0, & TO]) & 0.002 & ([0.002, & TO])\\
45-10-p0.75-hard-y & 100 & 263 & ([248, & 299]) & TO & ([3.333, & TO]) & 8.1615 & ([1.414, & 58.377]) & 91.5115 & ([6.605, & 215.261]) & 98.7765 & ([6.156, & TO]) & 1.2565 & ([0.093, & 2.906])\\
45-10-p0.75-hard-n & 100 & 263 & ([248, & 299]) & TO & ([TO, & TO]) & TO & ([277.753, & TO]) & TO & ([TO, & TO]) & 205.855 & ([47.682, & TO]) & 2.782 & ([2.003, & 5.963])\\
\end{tabular}

    \caption{Performance of various solvers for Dominating $M_3$-Detection. For each solver, we report median times [minimum time, maximum time] in seconds over the benchmark. 'TO' indicates the timeout value of 5 min. A value of 0 denotes times less than 1ms. \label{tab:M3}}
\end{table}

\begin{table}
\footnotesize
\begin{tabular}{l|rr@{$\;$}r@{}rr@{$\;$}r@{}r|r@{$\;$}r@{}r|r@{$\;$}r@{}rr@{$\;$}r@{}r|r@{$\;$}r@{}rr@{$\;$}r@{}r}
        Dataset & Size & \multicolumn{3}{c}{$n$} & \multicolumn{3}{c}{$m$} & \multicolumn{3}{c}{eGss} & \multicolumn{3}{c}{kissat}  & \multicolumn{3}{c}{Gurobi} & \multicolumn{3}{c}{proto-BnB} & \multicolumn{3}{c}{BnB-Cycle}\\
        \hline
rome-y & 25 & 12 & ([10, & 17]) & 18 & ([12, & 31]) & 0 & ([0, & 0]) & 0.001 & ([0.001, & 0.025]) & 0 & ([0, & 0]) & 0 & ([0, & 0]) & 0 & ([0, & 0])\\
rome-n & 11504 & 47 & ([10, & 107]) & 63 & ([9, & 158]) & 0 & ([0, & 0.003]) & 0.038 & ([0.001, & 0.106]) & 0.136 & ([0.002, & 1.851]) & 0 & ([0, & 0]) & 0 & ([0, & 0])\\
HoG-ds...-y & 635 & 28 & ([20, & 171]) & 103 & ([60, & 9435]) & 0 & ([0, & 0.01]) & 0.002 & ([0.001, & 0.13]) & 0.029 & ([0, & 1.464]) & 0 & ([0, & 0]) & 0 & ([0, & 0.006])\\
HoG-ds...-n & 764 & 36 & ([20, & 162]) & 199 & ([60, & 4536]) & 0.012 & ([0.001, & 100.524]) & 0.079 & ([0.001, & 141.591]) & 0.9215 & ([0.003, & 113.874]) & 0 & ([0, & 1.926]) & 0.002 & ([0, & 16.981])\\
stride-y & 120 & 200 & ([100, & 1049]) & 8028.5 & ([243, & 93800]) & 0.0165 & ([0, & 6.839]) & 0.081 & ([0.006, & 31.679]) & 0.3235 & ([0.014, & 263.442]) & 0 & ([0, & 0.85]) & 0.0055 & ([0, & 103.531])\\
stride-n & 128 & 179.5 & ([100, & 1329]) & 2331.5 & ([196, & 46627]) & 0.048 & ([0.002, & 125.491]) & 0.29 & ([0.062, & 145.055]) & 4.781 & ([0.204, & 293.427]) & 0 & ([0, & 53.879]) & 0 & ([0, & 28.562])\\
\end{tabular}

    \caption{Performance of various solvers for Dominating $C_5$-Detection. For each solver, we report median times [minimum time, maximum time] in seconds over the benchmark. 'TO' indicates the timeout value of 5 min. A value of 0 denotes times less than 1ms. \label{tab:C5}}
\end{table}

\begin{table}
\small \setlength{\tabcolsep}{3pt}
\begin{tabular}{l|rr@{$\;$}r@{}rr@{$\;$}r@{}r|r@{$\;$}r@{}r|r@{$\;$}r@{}rr@{$\;$}r@{}r|r@{$\;$}r@{}rr@{$\;$}r@{}r}
        Dataset & Size & \multicolumn{3}{c}{$n$} & \multicolumn{3}{c}{$m$} & \multicolumn{3}{c}{eGSS} & \multicolumn{3}{c}{kissat}  & \multicolumn{3}{c}{Gurobi} & \multicolumn{3}{c}{proto-BnB} & \multicolumn{3}{c}{BnB-Path}\\
        \hline
rome-y & 212 & 12 & ([10, & 17]) & 16 & ([10, & 31]) & 0 & ([0, & 0.001]) & 0.002 & ([0.001, & 0.009]) & 0.003 & ([0, & 0.035]) & 0 & ([0, & 0]) & 0 & ([0, & 0])\\
rome-n & 11320 & 47 & ([10, & 110]) & 64 & ([9, & 158]) & 0 & ([0, & 0.008]) & 0.044 & ([0.001, & 0.132]) & 0.111 & ([0.002, & 1.21]) & 0 & ([0, & 0]) & 0 & ([0, & 0])\\
HoG-ds...-y & 639 & 26 & ([20, & 171]) & 101 & ([60, & 9435]) & 0 & ([0, & 0.176]) & 0.002 & ([0.001, & 1.193]) & 0.009 & ([0, & 3.542]) & 0 & ([0, & 0.015]) & 0 & ([0, & 0.255])\\
HoG-ds...-n & 760 & 34 & ([20, & 162]) & 160 & ([60, & 4536]) & 0.022 & ([0.001, & 102.221]) & 0.112 & ([0.001, & TO]) & 1.155 & ([0.003, & TO]) & 0.001 & ([0, & 6.174]) & 0.004 & ([0, & 59.178])\\
stride-y & 169 & 188 & ([100, & 1049]) & 6502 & ([243, & 93800]) & 0.019 & ([0, & 79.141]) & 0.215 & ([0.007, & 103.148]) & 0.445 & ([0.01, & TO]) & 0 & ([0, & 2.47]) & 0.002 & ([0, & 67.068])\\
stride-n & 78 & 179.5 & ([100, & 1329]) & 2498 & ([196, & 40411]) & 0.0645 & ([0.004, & TO]) & 0.578 & ([0.061, & TO]) & 6.9435 & ([0.181, & TO]) & 0.0015 & ([0, & 212.471]) & 0.001 & ([0, & 152.943])\\
\end{tabular}

    \caption{Performance of various solvers for Dominating $P_5$-Detection. For each solver, we report median times [minimum time, maximum time] in seconds over the benchmark. 'TO' indicates the timeout value of 5 min. A value of 0 denotes times less than 1ms. \label{tab:P5}}
\end{table}

    \end{landscape}
    \clearpage

\begin{figure*}[t]
\begin{subfigure}[t]{0.5\textwidth}
\includegraphics[width=\textwidth]{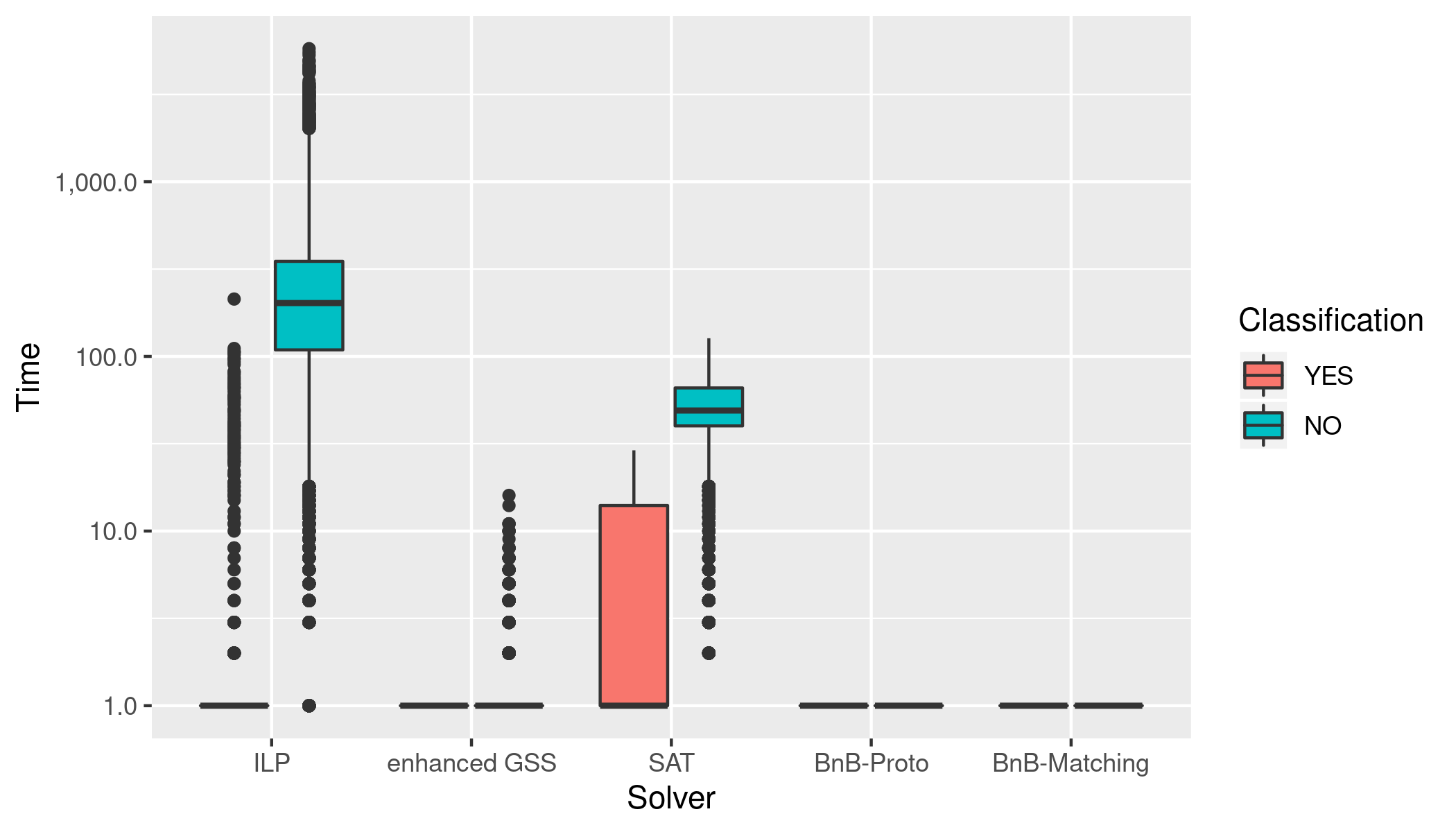}
\caption{Rome}
\end{subfigure}
\begin{subfigure}[t]{0.5\textwidth}
\includegraphics[width=\textwidth]{figures/M3-on-HoG-ds1-10-hard.png}
\caption{HoG-ds1-10-hard}
\end{subfigure}
\begin{subfigure}[t]{0.5\textwidth}
\includegraphics[width=\textwidth]{figures/M3-on-stride-package1.png}
\caption{stride}
\end{subfigure}
\begin{subfigure}[t]{0.5\textwidth}
\includegraphics[width=\textwidth]{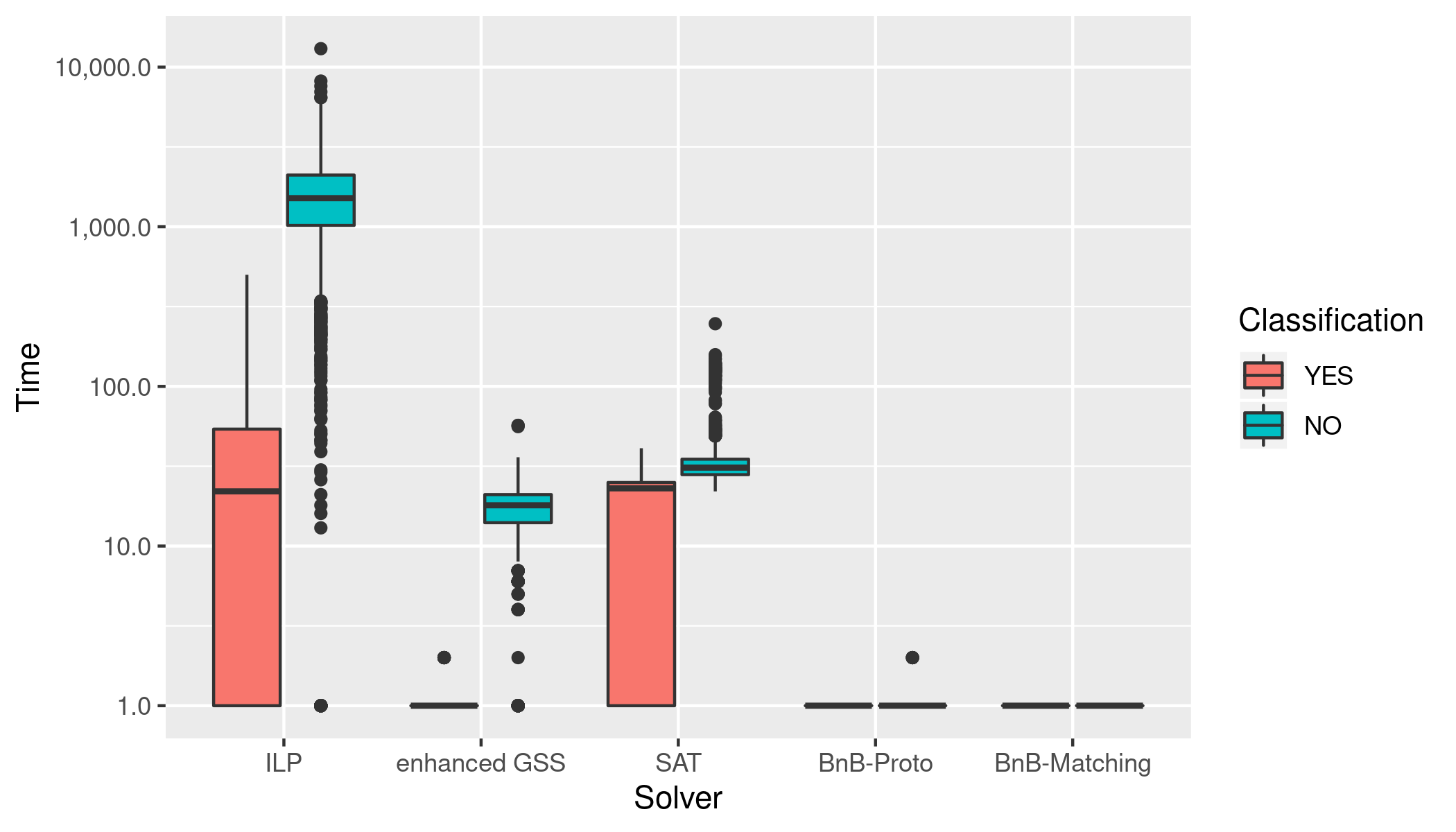}
\caption{HoG-M3}
\end{subfigure}
\begin{subfigure}[t]{0.5\textwidth}
\includegraphics[width=\textwidth]{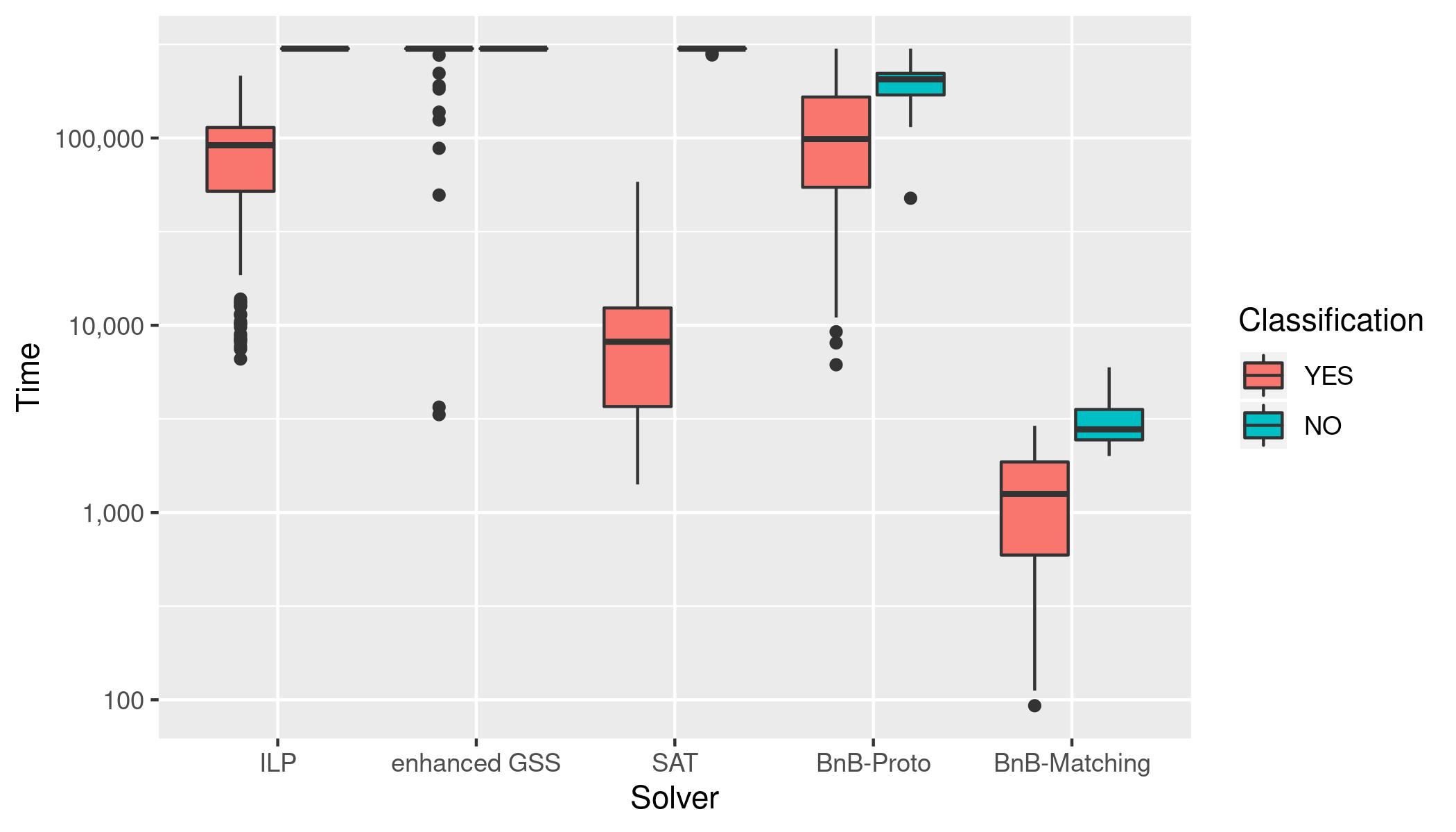}
\caption{OV-M3}
\end{subfigure}
\caption{Running times for Dominating $M_3$-Pattern on different datasets in ms\label{fig:appendix-M3}}
\end{figure*}

\begin{figure*}[t]
\begin{subfigure}[t]{0.5\textwidth}
\includegraphics[width=\textwidth]{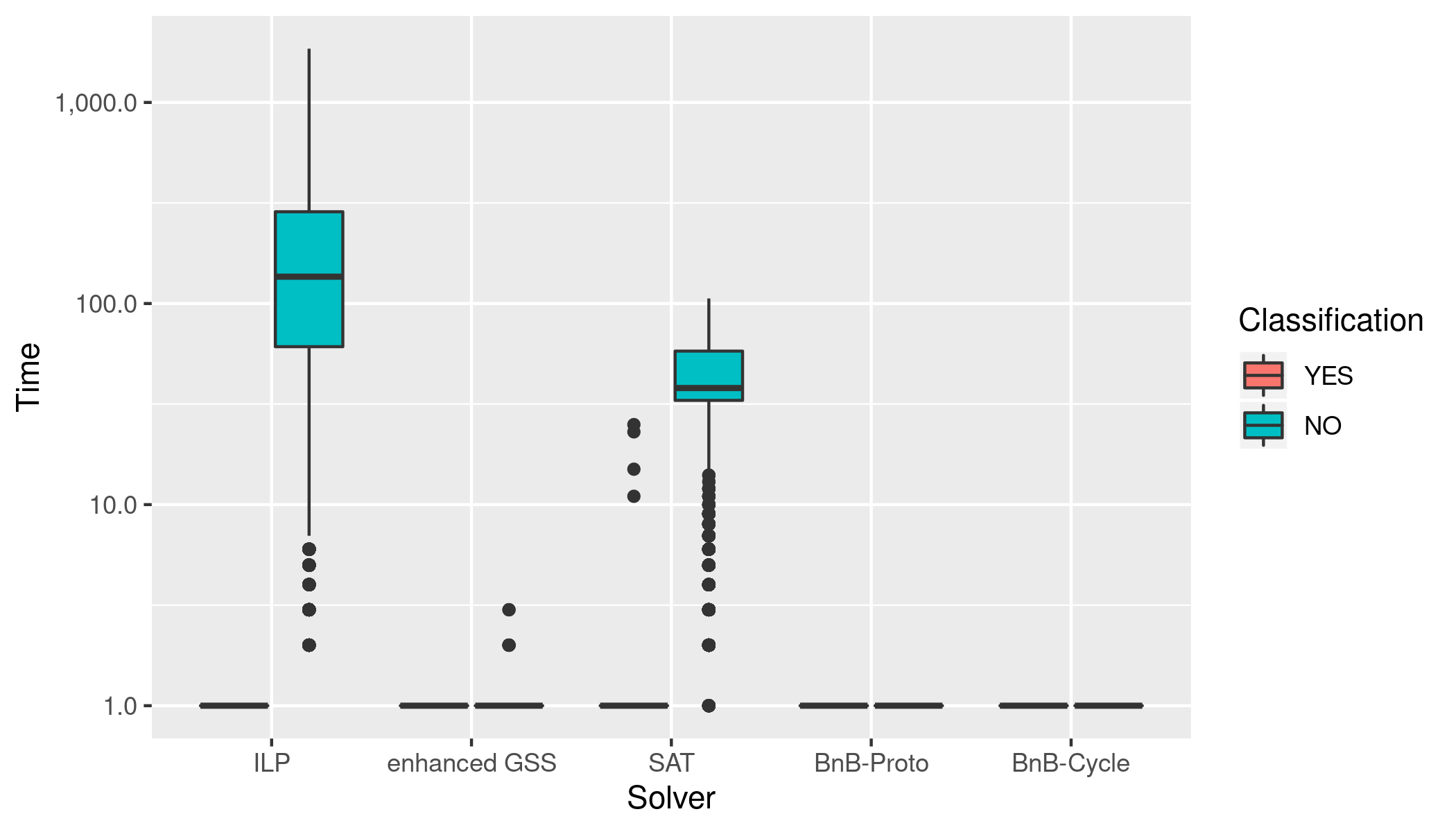}
\caption{Rome}
\end{subfigure}
\begin{subfigure}[t]{0.5\textwidth}
\includegraphics[width=\textwidth]{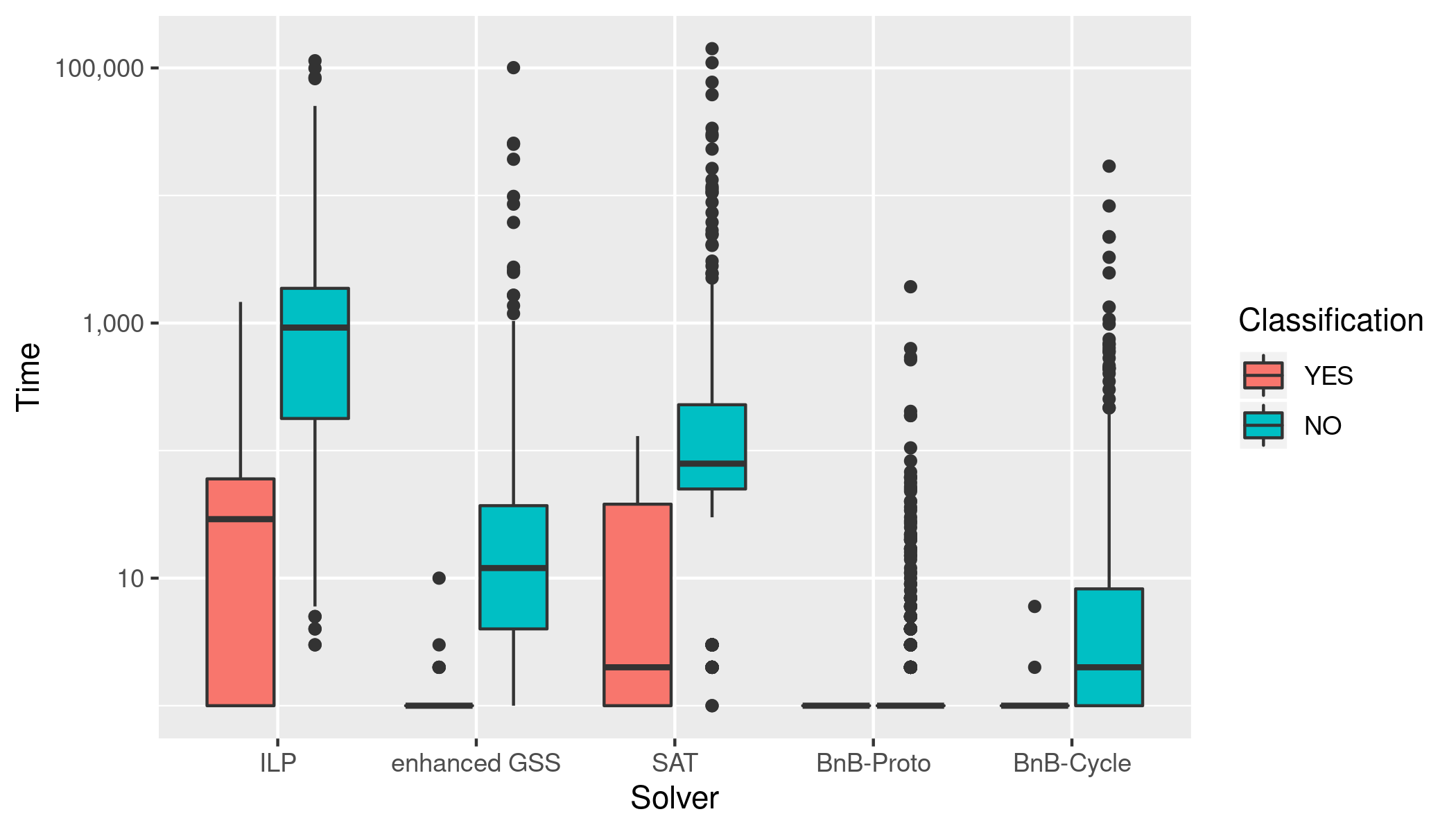}
\caption{HoG-ds1-10-hard}
\end{subfigure}
\begin{subfigure}[t]{0.5\textwidth}
\includegraphics[width=\textwidth]{figures/C5-on-stride-package1.png}
\caption{stride}
\end{subfigure}
%
\caption{Running times for Dominating $C_5$-Pattern on different datasets in ms\label{fig:appendix-C5}}
\end{figure*}

\begin{figure*}[t]
\begin{subfigure}[t]{0.5\textwidth}
\includegraphics[width=\textwidth]{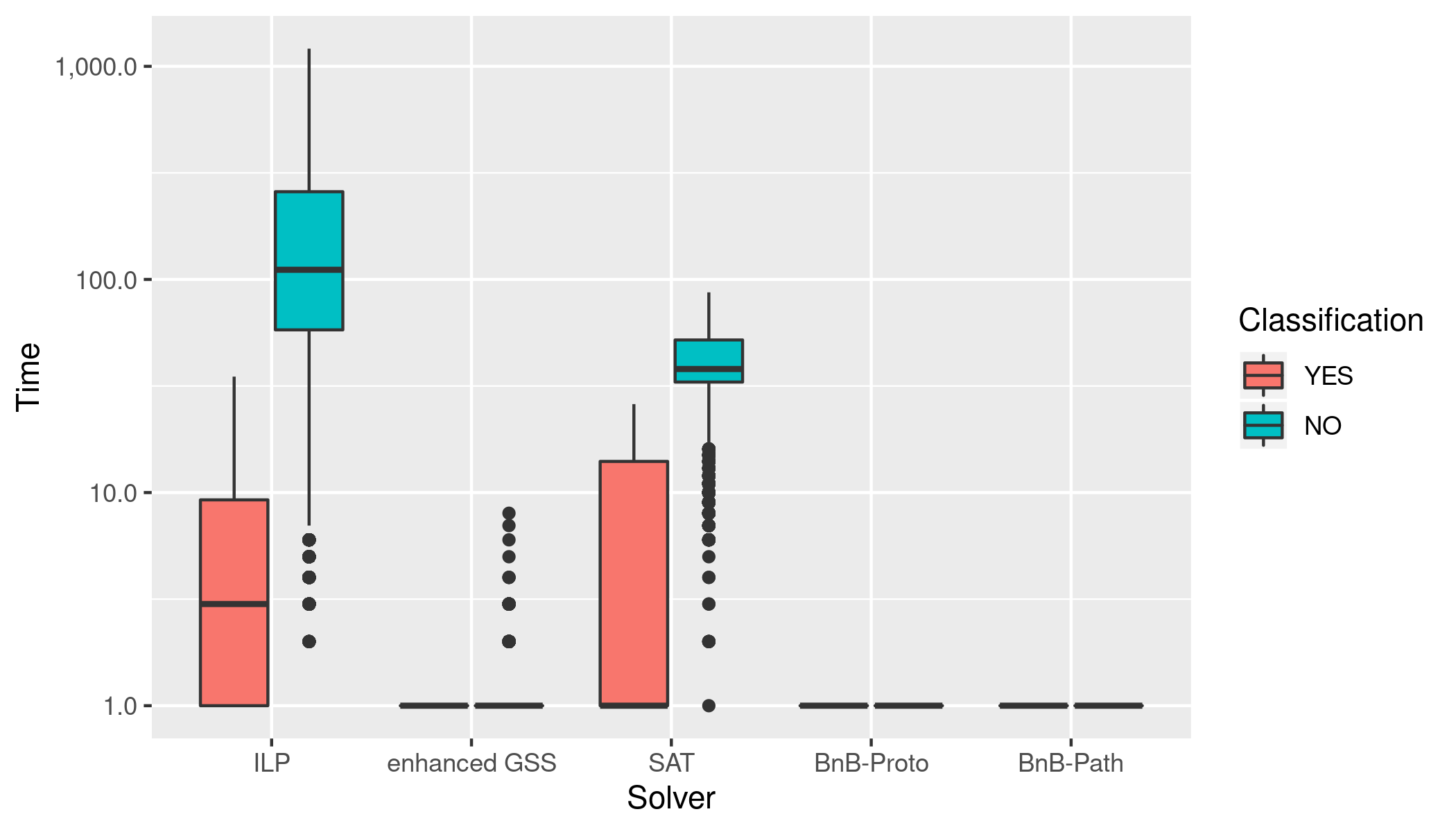}
\caption{Rome}
\end{subfigure}
\begin{subfigure}[t]{0.5\textwidth}
\includegraphics[width=\textwidth]{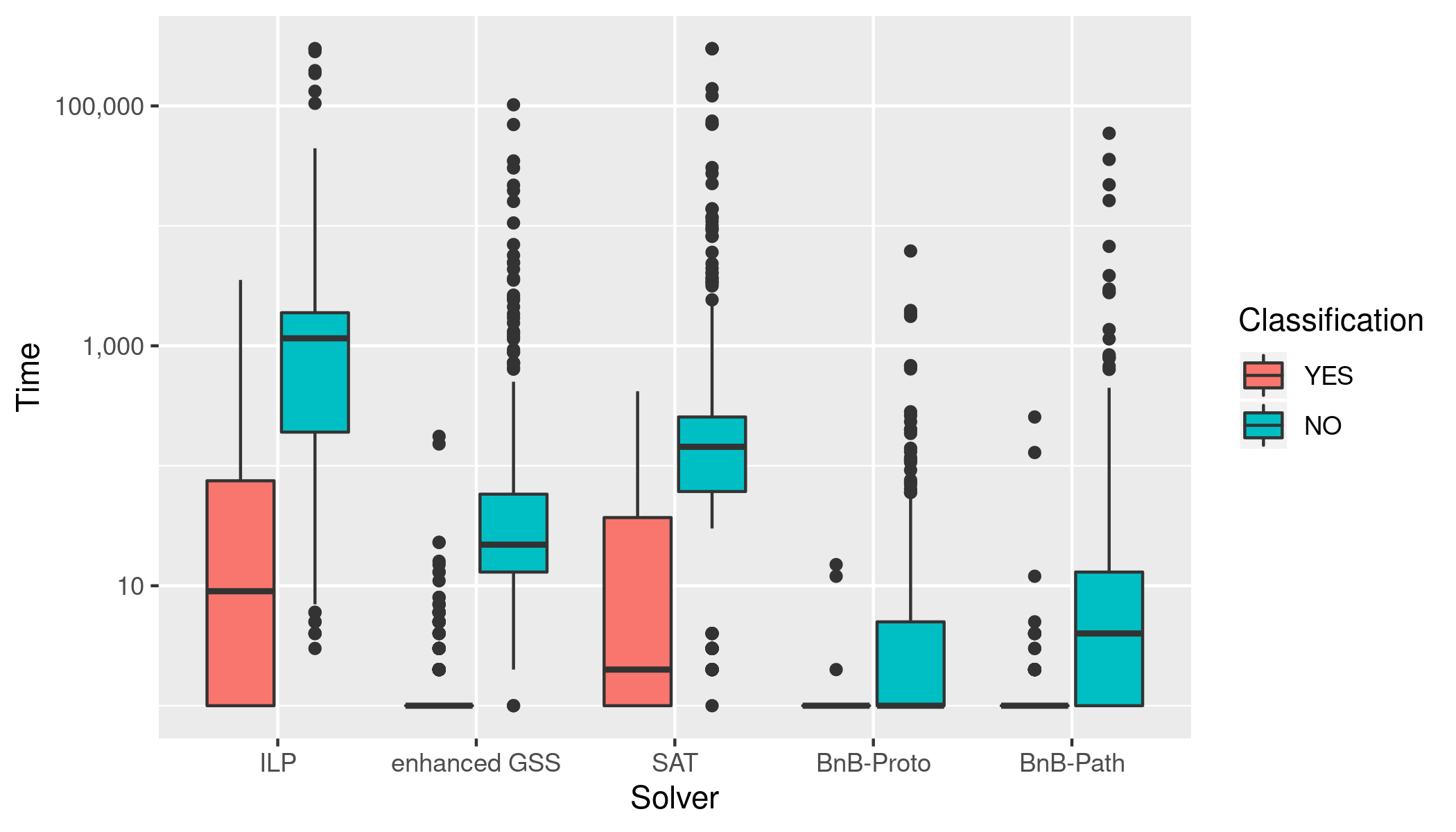}
\caption{HoG-ds1-10-hard}
\end{subfigure}
\begin{subfigure}[t]{0.5\textwidth}
\includegraphics[width=\textwidth]{figures/P5-on-stride-package1.png}
\caption{stride}
\end{subfigure}

\caption{Running times for Dominating $P_5$-Pattern on different datasets in ms\label{fig:appendix-P5}}
\end{figure*}

\end{document}